\documentclass[twocolumn]{aastex631}

\usepackage{amssymb}
\usepackage{amsmath}
\usepackage{comment}
\usepackage[colorlinks=true]{hyperref}
\usepackage{cleveref}

\shorttitle{Principle of Inhomogeneity I}
\shortauthors{Zhang}
\graphicspath{{./}{figures/}}
\begin{document}

\title{The Inhomogeneity Effect I:\\
Inhomogeneous Surface and Atmosphere Accelerate Planetary Cooling}

\correspondingauthor{Xi Zhang}
\email{xiz@ucsc.edu}

\author{Xi Zhang}
\affiliation{Department of Earth and Planetary Sciences, University of California Santa Cruz, Santa Cruz, CA 95064, USA}

%% Mark off the abstract in the ``abstract'' environment. 
\begin{abstract}

We propose a general principle that under the radiative-convective equilibrium, the spatial and temporal variations in a planet's surface and atmosphere tend to increase its cooling. This principle is based on Jensen's inequality and the curvature of the response functions of surface temperature and outgoing cooling flux to changes in incoming stellar flux and atmospheric opacity. We use an analytical model to demonstrate that this principle holds for various planet types: (1) on an airless planet, the mean surface temperature is lower than its equilibrium temperature; (2) on terrestrial planets with atmospheres, the inhomogeneity of incoming stellar flux and atmospheric opacity reduces the mean surface temperature; (3) on giant planets, inhomogeneously distributed stellar flux and atmospheric opacity increase the outgoing infrared flux, cooling the interior. Although the inhomogeneity of visible opacity might sometimes heat the atmosphere, the effect is generally much smaller than the inhomogeneous cooling effect of infrared opacity. Compared with the homogeneous case, the mean surface temperature on inhomogeneous terrestrial planets can decrease by more than 20\%, and the internal heat flux on giant planets can increase by over an order of magnitude. Despite simplifications in our analytical framework, the effect of stellar flux inhomogeneity appears to be robust, while further research is needed to fully understand the effects of opacity inhomogeneity in more realistic situations. This principle impacts our understanding of planetary habitability and the evolution of giant planets using low-resolution and one-dimensional frameworks that may have previously overlooked the role of inhomogeneity. 
        
\end{abstract}

%% Keywords should appear after the \end{abstract} command. 
%% The AAS Journals now uses Unified Astronomy Thesaurus concepts:
%% https://astrothesaurus.org
%% You will be asked to selected these concepts during the submission process
%% but this old "keyword" functionality is maintained in case authors want
%% to include these concepts in their preprints.

%% From the front matter, we move on to the body of the paper.
%% Sections are demarcated by \section and \subsection, respectively.
%% Observe the use of the LaTeX \label
%% command after the \subsection to give a symbolic KEY to the
%% subsection for cross-referencing in a \ref command.
%% You can use LaTeX's \ref and \label commands to keep track of
%% cross-references to sections, equations, tables, and figures.
%% That way, if you change the order of any elements, LaTeX will
%% automatically renumber them.
%%
%% We recommend that authors also use the natbib \citep
%% and \citet commands to identify citations.  The citations are
%% tied to the reference list via symbolic KEYs. The KEY corresponds
%% to the KEY in the \bibitem in the reference list below. 

\section{Introduction} \label{introsec}

The surface and atmosphere of a three-dimensional (3D) planet are characterized by spatial inhomogeneity and temporal variability, which can arise from various sources. The incident stellar flux on a spherical planet is spatially inhomogeneous due to the incident angle. It may also exhibit complex spatial and temporal variations depending on the planet's rotation, obliquity, and orbital eccentricity. An extreme example is the tidally locked exoplanets with permanent dayside and nightside. Inhomogeneities in the surface may also be caused by variations in thermal inertia, albedo, and emissivity. The presence of an atmosphere further complicates the situation, as surface-atmosphere interactions can redistribute materials such as water, sand, and ice on the surface.

Physical and chemical processes in the atmosphere often induce significant inhomogeneities. Short-lived chemical tracers are known to display local variations (\citealt{zhangGlobalmeanVerticalTracer2018,zhangGlobalmeanVerticalTracer2018a}). The ubiquity of patchy clouds on planets (\citealt{zhangStratosphericAerosolsJupiter2013, parmentier3DMixingHot2013,powellTransitSignaturesInhomogeneous2019,feinsteinEarlyReleaseScience2022}) alters planetary albedo and infrared opacity. Furthermore, atmospheric dynamics contribute to the redistribution of heat and chemical species (\citealt{geRotationalLightCurves2019,drummondImplicationsThreedimensionalChemical2020,gilliVenusUpperAtmosphere2021,shaoLocaltimeDependenceChemical2022,leeMiniChemicalSchemeNet2023}). On irradiated planets, the spatial variation of incoming stellar flux produces not only temperature contrasts, but also uneven distributions of photochemically produced species (\citealt{agundezPseudo2DChemical2014,shaoLocaltimeDependenceChemical2022}). For free-floating planets or brown dwarfs, convective storms, vortices, waves, jets, and cloud condensation can manifest significant surface inhomogeneities, resulting in rotational modulations (\citealt{artigauPhotometricVariabilityT22009,billerVariabilityYoungTransition2015,zhouDiscoveryRotationalModulations2016,apaiZonesSpotsPlanetaryscale2017,zhangAtmosphericCirculationBrown2014,tanAtmosphericCirculationBrown2021a,tanAtmosphericCirculationBrown2021}).

It has been shown that atmospheric inhomogeneity is essential to allow the upward mixing of chemical tracers by the large-scale circulation (\citealt{zhangGlobalmeanVerticalTracer2018, zhangGlobalmeanVerticalTracer2018a}). However, its effect on energy transport has not been thoroughly examined. As the radiative transport processes depend critically on temperature and opacity, it is expected that planetary inhomogeneity may influence incoming and outgoing energy fluxes and regulate surface temperatures on terrestrial planets, as well as the interior cooling rate on giant planets. To date, one-dimensional (1D) atmospheric models have been widely used to estimate the mean surface temperature of habitable planets (e.g., \citealt{kastingHabitableZonesMain1993, kopparapuHABITABLEZONESMAINSEQUENCE2013}) and calculate energy flux for the interior evolution of giant planets (e.g., \citealt{hubbardJovianSurfaceCondition1977, guillotGiantPlanetsSmall1996, fortneyPlanetaryRadiiFive2007}). In these models, the incoming and outgoing fluxes, temperature, and opacity sources are assumed to be uniformly distributed over the globe. While using 1D models is computationally efficient for long-time evolution calculations, the importance of inhomogeneity---a fundamental property of a realistic 3D planetary atmosphere---has not been systematically evaluated.

Several previous studies recognized the impact of surface and atmospheric inhomogeneity. Earth climate studies have examined the spatial heterogeneity of opacity sources such as clouds and highlighted the potential problems with using average values (e.g., \citealt{larsonSystematicBiasesMicrophysics2001, fauchezImpactCirrusClouds2014}). Detailed calculations demonstrated that regional climate feedback does affect the global climate forcing (e.g., \citealt{armourTimevaryingClimateSensitivity2013}). \cite{leconte3DClimateModeling2013} argued that equilibrium temperature is not an accurate measure of the mean surface temperature of terrestrial planets, not only due to the greenhouse effect but also the nonlinearity of blackbody emission, which touches upon the essence of the inhomogeneity discussed in this study. They further quantify the difference between the 1D and 3D atmospheric models for dry, synchronously rotating terrestrial planets and found that the global-mean surface temperature in 1D models is hotter than that in 3D models (see their Fig. 6). Because both models adopted the same stellar flux and radiative schemes, the difference is likely a result of the spatial inhomogeneity caused by the 3D circulation pattern. If the moisture is included, 3D climate simulations have shown that inhomogeneously distributed clouds can significantly alter surface temperature and planetary habitability (e.g., \citealt{yangStabilizingCloudFeedback2013, turbetDayNightCloud2021}). Moreover, temporal variability causes the mean temperature difference. For example, a simple energy balance model shows a lower diurnal-mean temperature than the flux-weighted mean temperature due to the nonlinearity of blackbody emission (e.g., \citealt{lohmannTemperaturesEnergyBalance2020}). 

For giant planets, atmospheric inhomogeneity can also significantly affect the cooling rate of the interior. \cite{guillotEvolution51Pegasus2002} argued that the day-night contrast increases the interior cooling of tidally locked gas giants and leads to a smaller planetary radius (also see \citealt{budajDayNightSide2012, spiegelThermalProcessesGoverning2013}). \cite{rauscherINFLUENCEDIFFERENTIALIRRADIATION2014} studied a zonally symmetric planet using a radiative model and found an inhomogeneous radiative-convective boundary (RCB) from the equator to the poles can enhance cooling compared to a uniform RCB on hot Jupiters.

The aforementioned studies on different types of planets point toward the same conclusion that the inhomogeneities of planetary atmosphere and surface significantly affect planetary cooling. To synthesize the thoughts in previous literature, here we conduct a systematic study of the effect of inhomogeneity on generic planets. Based on the conservation laws of energy and matter, we focus on the inhomogeneity of incident stellar flux and atmospheric opacity as the key factors influencing temperature structure and radiative flux to space. Using a radiative-convective equilibrium (RCE) analytical model with gray opacity and without scattering, we simulate the atmospheric structure, surface temperature, and outgoing infrared flux for terrestrial and giant planets for each atmospheric column. We then compare the results of simple non-uniform cases, which include two inhomogeneous columns, to the results of the uniform cases, which consist of a single column with an average incident stellar flux and opacity. As a first study of the effect of inhomogeneity on planetary heat flow, an analytical approach allows for clearer physical insights. Additionally, it can minimize numerical errors when solving the radiative transfer equation in regions of large opacity.

Our findings suggest that inhomogeneity in a planet's surface or atmosphere generally leads to faster cooling than a homogenous counterpart. This principle appears to hold across a wide range of parameters in our analytical framework, assuming RCE. On terrestrial planets, inhomogeneity in the surface or atmosphere leads to a reduction in mean surface temperature, while on giant planets, atmospheric inhomogeneity enhances interior cooling. Further numerical research can quantify the impact of inhomogeneity on planetary cooling more accurately beyond our analytical RCE framework, considering more realistic treatments of atmospheric opacity and dynamics.

We present our results in three consecutive papers in increasing order of complexity. In the first paper (the current one), we introduce the nature of the problem and the mathematical and physical foundation of the theory of the inhomogeneity effect. We then apply it to several planet types by comparing inhomogeneous and homogeneous cases. In the second paper (\citealt{zhangInhomogeneityEffectII2023}), we will further develop the theory to allow comparison among inhomogeneous cases, allowing us to quantify the dependence of planetary cooling on the degree of inhomogeneity. We will also examine how rotational and orbital configurations impact the global surface cooling of airless planets and the interior cooling of giant planets. In the third paper (\citealt{zhangInhomogeneityEffectIII2023}), we will specifically focus on tidally locked giant planets, using a non-hydrostatic general circulation model to explore how atmospheric inhomogeneity affects their interior cooling and our understanding of the heating mechanisms on these planets.

In this paper, we first introduce a simple version of the inhomogeneity effect theory in Section \ref{version0} and demonstrate how it is based on the convexity of the Planck function and Jensen's inequality in statistics. We then apply it to an airless, rocky body to show the effect of surface inhomogeneity. In Section \ref{rcesec}, we present our analytical RCE framework, focusing on evaluating the position of the RCB. We use the theory and model to explore the effect of atmospheric inhomogeneity on surface temperature for terrestrial planets in Section \ref{tesec}. We then apply the theory to giant planets to examine the effect of atmospheric inhomogeneity on outgoing infrared flux at the top of the atmosphere in Section \ref{gsec}. Finally, in Section \ref{consec}, we summarize the theory and its implications for our understanding of planetary cooling and discuss the limitations of our analytical framework.

\section{Version 0: Inhomogeneous brightness temperature enhances planetary cooling}\label{version0}

\subsection{Convexity of the Planck Function and Jensen's Inequality}

Before delving into specific physical processes that result in inhomogeneity on the surface and in the atmosphere of a planet, we first provide a simplified overview of the concept. Let us consider a generic planet with a non-uniform distribution of brightness temperature, which is a direct measure of the outgoing infrared radiation, whether or not the planet has an atmosphere. The average outgoing emission flux $F_\nu$ at frequency $\nu$ can be expressed as using brightness temperature $T_{\rm B}$:

\begin{equation}
\overline{F_\nu}=\pi \overline{B_\nu (T_{\rm B})},
\end{equation}
where $T_{\rm B}$ is the local brightness temperature and $B_\nu$ is the Planck function. The spatial and temporal average of the radiance is represented by $\overline{(\cdot)}$.

In low-resolution or one-dimensional models, one commonly calculates the outgoing flux using the averaged brightness temperature via $\pi B_\nu\left(\overline{T_{\rm B}}\right)$. We found that an inhomogeneous planet should emit more infrared flux than one with a homogeneous planet with the averaged temperature at every frequency:
\begin{equation}
\overline{B_\nu (T_{\rm B})} \ge B_\nu\left(\overline{T_{\rm B}}\right).\label{eq:v0}
\end{equation}
In other words, the entire spectrum is elevated with inhomogeneously distributed brightness temperature. The equality holds only when the brightness temperature is distributed uniformly. Another way to express this principle is that if two planets emit the same averaged flux, the uniform planet should have a higher global-mean brightness temperature than the non-uniform planet. 

\begin{figure}
\centering \includegraphics[width=0.48\textwidth]{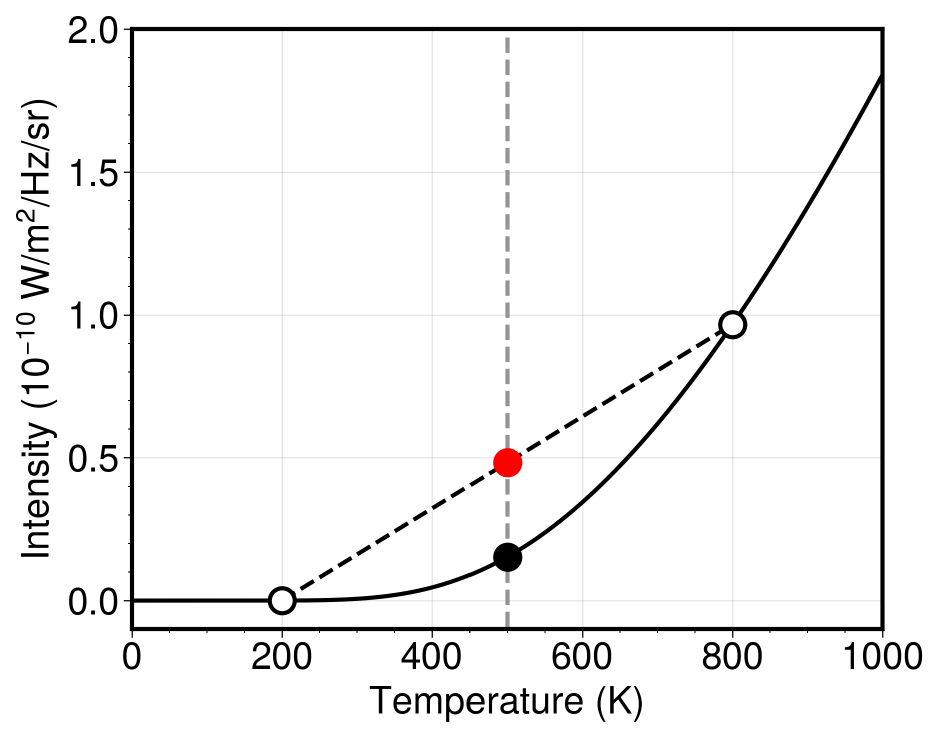} 
  \caption{Planck blackbody emission with brightness temperature at a frequency of 50 THz ($\sim6~\mu m$). The black dots represent the Planck emission at 200 K, 800 K, and intensity at their average temperature of 500 K (black filled dot), which is smaller than the average intensity (red filled dot) of the 200 K and 800 K emissions. The inequality arises from the convexity of the Planck function.} \label{fig:tplanck}
\end{figure}

 \Cref{eq:v0} stems from the fact that the Planck function $B_\nu$ is a convex function of temperature at any frequency. As we are unaware of any prior demonstrations of this fact, we have included a proof in Appendix \ref{app:planck}. Graphically speaking, the convex nature of $B_\nu$ implies that a linear combination of two brightness temperatures would yield a lower outgoing flux, as shown in \Cref{fig:tplanck}. This process can be repeated for all brightness temperatures over the globe to obtain the global mean and naturally result in the inequality in \Cref{eq:v0}. 

To provide an intuitive understanding, consider the gray limit scenario, where all frequencies share the same brightness temperatures. In this case, integrating the Planck function across frequency results in the Stefan-Boltzmann law $F=\sigma T_{\rm B}^4$, where $\sigma$ denotes the Stefan-Boltzmann constant. Clearly, this function is strictly convex in relation to temperature. \Cref{eq:v0} extends the convexity of this grey limit to non-gray cases. 

The mathematical foundation of \Cref{eq:v0} is based on Jensen's inequality (\citealt{jensenFonctionsConvexesInegalites1906}) which was also proved earlier by \cite{holder_1889} for doubly-differentiable functions. The finite form of Jensen's inequality states that for a convex function $f(x)$ and a set of real numbers $X = {x_1, x_2, ..., x_n}$, and a set of positive weights $W = {w_1, w_2, ..., w_n}$ such that the sum of the weights is equal to 1 (i.e., $\sum_{i=1}^{n} w_i = 1$), the following inequality holds:
\begin{equation}
\sum_{i=1}^{n} w_i f(x_i) \ge f\left(\sum_{i=1}^{n} w_i x_i\right) . \label{jineq}
\end{equation}
If we let all weights be equal $w_i=\frac{1}{n}$, the weighted averages are the mean values. It can also be generalized to a continuous form. Using the language of statistics, we can also state the inequality as:
\begin{equation}
\mathbb{E}\left[f(x)\right] \ge f\left(\mathbb{E}[x]\right).
\end{equation}
For any convex function, the expected value ($\mathbb{E}$) of the function ($f$) is greater than the function of the expected value. Therefore, for the convex function $B_\nu(T_{\rm B})$, we obtain \Cref{eq:v0}.

For a simpler Stefan-Boltzmann function $F=\sigma T_{\rm B}^4$, one can also apply the Roger-H{\"o}lder inequality (\citealt{rogersExtensionCertainTheorem1888,holder_1889})---which can be derived from the Jensen's equality---to the global average over the surface $S$:
\begin{equation}
\frac{1}{S}\int T_{\rm B} dS \le \left(\frac{1}{S}\int T_{\rm B}^4 dS\right)^{1/4}.
\end{equation}
Thus, the global mean brightness temperature of a non-uniform planet should be lower than that of a uniform planet. This result has been established in previous studies (e.g., \citealt{leconte3DClimateModeling2013,lohmannTemperaturesEnergyBalance2020}) using temperature instead of brightness temperature.

The foregoing discussion sets the basic philosophy for later evaluations of the non-linear responses of cooling flux and surface temperature to different parameters. The convex or concave characteristics and the application of Jensen's inequality are critical for comprehending the impact of inhomogeneity on planetary cooling. While it may not always be feasible to analytically demonstrate the convex nature of every response function, numerical evaluation can still offer meaningful insights into the system's general characteristics.

\subsection{Application to An Airless Body}

Here we provide a more realistic example with an airless body. The surface temperature is primarily determined by the incoming stellar flux and surface properties. For simplicity, we assume that the surface properties---such as the thermal inertia, albedo, and emissivity---are uniformly distributed on the planet as a zeroth-order assumption for many exoplanets. The temperature inhomogeneity primarily arises from planetary rotation and orbital configurations. The local surface temperature of a planet can be homogenized by planetary rotation. But even with the fastest planetary rotation, the surface temperature can only be distributed uniformly with longitude. The temperature inhomogeneity still exists due to the latitudinal distribution of the stellar flux. An immediate conclusion is that the mean surface temperature of an airless, rotating planet should be lower than its equilibrium temperature, which assumes global energy balance but ignores the inhomogeneity effect. 

An extreme case is a tidally locked planet. \cite{leconte3DClimateModeling2013} calculated that the mean surface temperature is about 0.57 times its equilibrium temperature. The diurnal mean temperature from a simple energy balance model also shows a lower temperature than the flux-weighted mean temperature (e.g., ($\overline{T^4})^{1/4}$) by the same factor (\citealt{lohmannTemperaturesEnergyBalance2020}). The influence of spatial inhomogeneities of thermal inertia, albedo, and emissivity deserves more exploration in future work. Is the mean surface temperature on a slower-rotating (more inhomogeneous) planet lower than that on a faster-rotating (less inhomogeneous) one? The answer goes beyond Jensen's inequality as it requires comparing two inhomogeneous cases. We will discuss this situation in the second paper (\citealt{zhangInhomogeneityEffectII2023}).

\subsection{Application to Planets with Atmospheres}

For planets with atmospheres, atmospheric inhomogeneity plays a crucial role in controlling planetary cooling and will be the primary focus of the rest of this paper. Heat transfer in planetary atmospheres is not only controlled by radiation but by many dynamical processes such as convection, large-scale circulation, and wave mixing. To keep the solutions analytically traceable, we simplified the treatment of dynamics in this study and assumed the atmosphere is composed of multiple, inhomogeneous 1D columns in RCE. Within each column, the heat transfer is carried out vertically by radiation in the top radiative zone and convection in the bottom convective zone. 

We further considered two types of planets with atmospheres: ``terrestrial" planets and ``giant" planets. In the terrestrial case, the internal heat flux is negligible, but the atmosphere and underlying surface are strongly coupled in each column. Vertical columns remain independent, as the horizontal mixing in the convective zone might not be strong enough to homogenize the temperature across the pressure surface (see Figure 6 \citealt{leconte3DClimateModeling2013}). We focused on the mean surface temperature change to explore the inhomogeneity effect. 

On giant planets with deep convective atmospheres and internal heat fluxes, the radiative timescale in the deep convective zone is very long (\citealt{showmanNumericalSimulationsForced2007}). The entropy in the convective zone is likely uniform because the superadiabaticity required to transport the intrinsic heat is very small (\citealt{guillotInteriorJupiter2004}). In addition, the infrared observations from Pioneer 10 and 11 discovered a latitudinally uniform temperature distribution in the troposphere of Jupiter   (e.g., \citealt{ingersollPioneer11Infrared1975}). Hence, we assumed the thermal structure in all columns follows the same adiabatic below the RCB in the giant planet scenario. We specified the deep entropy and calculated the global cooling flux at the top of the atmosphere (i.e., the internal heat flux) to explore the inhomogeneity effect.

We mainly explore two sources of inhomogeneity in the atmosphere over the constant pressure surface (i.e., isobar): (1) the inhomogeneity from incident stellar flux due to the variations in the incident angle or orbital distance and (2) the opacity inhomogeneity due to horizontal transport or chemistry. We will first explore the nonlinear dependence of surface temperature and internal heat flux on the incident stellar flux and opacity. We then calculate the atmosphere with inhomogeneous columns and then averaged the incident stellar flux/opacity over the isobars to mimic the homogeneous column. By comparing the two cases, we can quantify the inhomogeneity effects. 

\section{General Framework of a Radiative-Convective Atmosphere}\label{rcesec}

Here we introduce a 1D analytical RCE framework to solve a generic radiative-convective atmosphere. 1D RCE models have a long history in atmospheric science (e.g.,\citealt{manabeThermalEquilibriumAtmosphere1964,  goodyAtmosphericRadiationTheoretical1995,  liouIntroductionAtmosphericRadiation2002}). They have also been extensively applied to exoplanets in the last three decades (e.g., \citealt{marleyAtmosphericEvolutionarySpectral1996a,  fortneyUnifiedTheoryAtmospheres2008}), including analytical or semi-analytical studies (e.g., \citealt{hubenyPossibleBifurcationAtmospheres2003,  hansenAbsorptionRedistributionEnergy2008,  guillotRadiativeEquilibriumIrradiated2010,  robinsonAnalyticRadiativeConvectiveModel2012, hengAnalyticalModelsExoplanetary2014, parmentierNongreyAnalyticalModel2014,   parmentierNongreyAnalyticalModel2015, tanWeakSeasonalityTemperate2022}). In the model, the atmosphere is divided into a top radiative zone and a bottom convective zone by the RCB. We first describe our treatments of each zone and then discuss how to find the RCB.

\subsection{The Convective Zone}

For convenience, in this study, we normalized the temperature and energy fluxes by a reference value $T_0$ and $4\sigma T_0^4$, respectively. One can take $T_0$ as the equilibrium temperature for instance. The vertical temperature profile in the convective zone below the RCB follows an adiabat. The normalized adiabat is given by 
\begin{equation}
T_{\rm conv}(p)=T_{\rm s}\left(\frac{p}{p_{\rm s}}\right)^{\frac{\zeta(\gamma-1)}{\gamma}}.
\end{equation}
On terrestrial planets, $T_{\rm s}$ is the surface temperature (here normalized by $T_0$) at the surface pressure $p_{\rm s}$.  On giant planets, $T_{\rm s}$ is the interior temperature (also normalized by $T_0$) at a reference pressure $p_{\rm s}$ in the deep convective region. $\gamma$ is the adiabatic index. $\zeta$ approximates moist adiabatic or non-adiabatic processes. A value of $\zeta=1$ corresponds to a dry adiabat. Moist convection processes can lead to much lower values of $\zeta$ ranging from 0.2 to 0.5 on Solar System planets (\citealt{robinsonAnalyticRadiativeConvectiveModel2012}).

As noted, we will explore the spatial inhomogeneity across the isobars. However, the radiative-convective solutions can be greatly simplified if we adopt the optical depth as the coordinates. To relate the pressure to the optical depth, here we assume that the infrared (IR) optical depth, $\tau$, monotonically changes with pressure in a power-law function, 
\begin{equation}
\tau= \kappa p^n=\tau_s\left(\frac{p}{p_{\rm s}}\right)^n,\label{tau}
\end{equation}
where $\kappa$ is a constant coefficient. The exponent $n$ is approximately unity for Doppler-broadened lines in the upper atmosphere and becomes larger than unity in the deep atmosphere due to pressure-broadened lines and collisional-induced absorptions. For example, for a solar-metallicity hydrogen-dominated atmosphere at about 2000 K, $n$ is estimated to be around 1.5 according to \cite{freedmanGaseousMeanOpacities2014}. The adiabatic temperature structure in the convective region can be expressed as follows: 
\begin{equation}
T_{\rm conv}^4(\tau)=T_{\rm s}^4\left(\frac{\tau}{\tau_{\rm s}}\right)^\beta\label{adt},
\end{equation}
where the ``adiabatic slope parameter" $\beta=4\zeta(\gamma-1)/n\gamma$ is assumed to be a constant in this study. It controls the slope of the adiabatic temperature profile in the deep convective zone. Our definition of $\beta$ is the same as that in \citet{ginzburgHotJupiterInflationDue2015} but differs from that in \citet{robinsonAnalyticRadiativeConvectiveModel2012}. The latter defines $\beta=\zeta(\gamma-1)/\gamma$. Our $\beta$ equals to their $4\beta/n$.

The value of $\beta$ depends on the adiabatic index $\gamma$, which relies on the atmospheric composition and temperature. $\gamma$ is 5/3 for a monatomic gas like helium or argon. For diatomic gases, $\gamma$ is around 7/5 at low temperatures but can approach 9/7 at high temperatures if all vibrational modes are excited (\citealt{zhangEffectsBulkComposition2017}). For more complex molecules, such as $\mathrm{CO_2}$, $\gamma$ is smaller, around 1.3. Using the example of $n=1.5$ and $\zeta=1$, the value of $\beta$ would be approximately 0.76 for $\gamma=7/5$ and is slightly larger than 1 for $\gamma=5/3$. See more discussion of $\beta$ in \citet{ginzburgHotJupiterInflationDue2015}.

\subsection{The Radiative Zone}\label{sec:rad}

In the radiative zone, we seek a temperature solution in radiative equilibrium. A semi-gray radiative transfer problem is considered with two spectral bands: stellar flux absorption in the visible band and self-emission in the thermal IR band. In Appendix \ref{app1}, we derived a new two-stream radiative diffusive equation in the no-scattering limit for a plane parallel atmosphere. The normalized form for the IR band is given by:
\begin{equation}
\frac{d^2F}{d\tau^2}=D^2F-\frac{dT_{\rm rad}^4}{d\tau}, 
        \label{trt}
\end{equation}
where $\tau$ is the infrared optical depth, $T_{\rm rad}$ is the normalized temperature in the radiative zone by $T_0$, and $F$ is the normalized net thermal infrared flux by $4\sigma T_0^4$.

\Cref{trt} is slightly different from the traditional diffusive equation. The diffusivity factor, $D$, is approximated differently in this theory as $D\approx E_1(x)/E_2(x) \approx E_2(x)/E_3(x)$ where $E_n(x)$ is the exponential integral function. In practice, $D$ is taken as a constant. This equation is proposed to unify the widely used forms under Eddington approximation and hemi-isotropic approximation. It converges to the equation under the Eddington approximation if $D=\sqrt{3}$ (e.g., \citealt{hubenyPossibleBifurcationAtmospheres2003, guillotRadiativeEquilibriumIrradiated2010}) or to the hemi-isotropic diffusive equation if $D=2$ (e.g., \citealt{robinsonAnalyticRadiativeConvectiveModel2012}). With this formalism, our results in this study do not depend on specific diffusive assumptions in radiative transfer. One can eliminate $D$ in the equation by scaling the optical depth and temperature $\tau^*=D\tau$ and $T^*_{\rm rad}=T_{\rm rad}D^{-1/4}$, respectively. Thus the behavior of the system is independent of the choice of $D$. In this work, we keep $D$ in the equation to maintain a clearer physical meaning of $\tau$ and $T_{\rm rad}$.

In radiative equilibrium (RE), the heating and cooling at every atmospheric level balance each other, and the total vertical flux are conserved:
\begin{equation}
\frac{\partial (F+F_{\rm v})}{\partial \tau}=0.
\end{equation}
$F_{\rm v}$ is the flux other than the IR radiative flux, such as the attenuated stellar flux in the visible band. The energy flux balance in the radiative zone is given by $F=-F_{\rm v}+F_{\rm int}$, where $F_{\rm int}$ is an integration constant determined by the boundary condition. Because the downward infrared flux vanishes at the top of the atmosphere, $F_{\rm int}$ is essentially the upward internal heat flux measured at the top, if there is any. 

Integrating \Cref{trt} gives the expression for the temperature $T_{\rm rad}$:
\begin{equation}
T_{\rm rad}^4(\tau)=-D^2\int F_{\rm v} d\tau+\frac{\partial F_{\rm v}}{\partial\tau}+D^2F_{\rm int}\tau+C.
\label{trad1}
\end{equation}
The constant $C$ can be determined using the radiative flux boundary condition at the top of the atmosphere. By using the upward ($F^+$) and downward ($F^-$) fluxes in \Cref{fluxeq}, we get:
\begin{equation}
\frac{dF}{d\tau}=-T_{\rm rad}^4 + D(F^+-F^-).
\end{equation}
At top of the atmosphere ($\tau=0$), there is no downward thermal radiative flux ($F^-=0$), so $F^+-F^-=F^++F^-=F=-F_{\rm v}+F_{\rm int}$. By using the temperature structure from \Cref{trad1}, we obtain $C$:
\begin{equation}
C=DF_{\rm int}+(D^2\int F_{\rm v} d\tau-DF_{\rm v})\Big|_{\tau=0}
\end{equation}

The radiative-equilibrium temperature structure is given by:
 \begin{equation}
    T_{\rm rad}^4(\tau)=S(\tau)+(D+D^2\tau)F_{\rm int}, \label{trad}
\end{equation}
where $S(\tau)$ is the contribution from $F_{\rm v}$:
 \begin{equation}
    S(\tau)=\frac{\partial F_{\rm v}}{\partial\tau}-D^2\int_0^\tau F_{\rm v}d\tau-DF_{\rm v}(0).\label{srceq}
\end{equation}
This is a general solution for an arbitrary form of $F_{\rm v}$. The associated fluxes $F^+_{\rm rad}(\tau)$ and $F^-_{\rm rad}(\tau)$ are derived in Appendix \ref{app2}. 

If $F_{\rm v}$ represents the attenuated stellar flux and follows Beer's law, we can write $F_{\rm v}=-F_{\odot}e^{-\alpha\tau/\mu}$, where $\alpha=\kappa_v/\kappa$ is the ratio of the visible opacity to the infrared, $\mu$ is the cosine of the incident angle, and $F_{\odot}$ is the local incident stellar flux normalized by $4\sigma T_0^4$. The temperature structure in the radiative zone using \Cref{trad} is given by:
\begin{equation}
\begin{split}
T_{\rm rad}^4(\tau)&=F_{\odot}\bigg[D+\frac{D^2\mu}{\alpha}+(\frac{\alpha}{\mu}-\frac{D^2\mu}{\alpha})e^{-\alpha\tau/\mu}\bigg]\\
&+ (D+D^2\tau)F_{\rm int}.
\end{split}
\end{equation}
This radiative equilibrium solution is consistent with that under the Eddington approximation if we adopt $D=\sqrt{3}$ (e.g., \citealt{hansenAbsorptionRedistributionEnergy2008, guillotRadiativeEquilibriumIrradiated2010}). For example, we can reproduce Equation (26) in \cite{guillotRadiativeEquilibriumIrradiated2010} with $f_{\rm Kth}=1/3$ and the second Eddington coefficient $f_{\rm Hth}=1/\sqrt{3}$. If we adopt $D=2$, our solution is consistent with the hemi-isotropic case. Additionally, by assuming two visible bands in $F_{\rm v}$, we can reproduce Equation (18) in \cite{robinsonAnalyticRadiativeConvectiveModel2012}. Using a free form of the $F_{\rm v}$ in our solution allows more flexibility to explore different heating mechanisms in the atmosphere.

\subsection{Onset of Convection}

To find out the RCB, it would be essential to discuss the criterion for the onset of convection. The Schwarzschild criterion has been widely used for a free atmosphere without surface interaction to analyze convective instability (e.g., \citealt{ginzburgExtendedHeatDeposition2016}). This criterion states that the atmospheric column is convectively unstable if the temperature gradient under radiative equilibrium exceeds the adiabatic lapse rate. As we have assumed that the atmospheric optical depth is a function of pressure but not temperature, we can write the criterion in the optical depth coordinate using \Cref{adt} and \Cref{trad}:
\begin{equation}
\frac{d\ln T_{\rm rad}^4}{d\ln \tau}\Big|_{\tau_{\rm rcb}}=\frac{d\ln T_{\rm conv}^4}{d\ln \tau}\Big|_{\tau_{\rm rcb}}=\beta.\label{sch}
\end{equation}
It is generally considered a sufficient condition for convective instability (i.e., the necessary condition for convective stability). 

For a deep, optically-thick atmosphere where the stellar flux is negligible ($F_{\rm v}\sim0$) but internal heat flux is important, the gradient of the radiative energy ($d\ln T_{\rm rad}^4/d\ln \tau$) reaches unity (Equation \ref{trad}). If the adiabatic slope parameter $\beta$ is larger than 1, the convective zone might not be developed (see more discussion in Appendix \ref{app3}). Assuming $n=1.5$ and $\zeta=1$, $\beta$ is usually smaller than 1 for the atmosphere composed of diatomic and more complex molecules. An interesting case is a helium-dominated atmosphere, which might result from severe fractionation during photoevaporation (\citealt{huHELIUMATMOSPHERESWARM2015,malskyCoupledThermalCompositional2020,malskyHeliumenhancedPlanetsUpper2022}). For monotonic gas, $\beta$ could be larger than 1, so a pure helium atmosphere might be stable against convection to the deep interior under this estimate. In reality, because hydrogen is still present, the deep atmosphere might still be convective, but the RCB would be located deeper than in a hydrogen-dominated atmosphere (e.g., \citealt{huHELIUMATMOSPHERESWARM2015}).

For a terrestrial planet with a relatively thin atmosphere and a surface but no interior heat flux, the convection criterion is different because the stellar flux could be important in the convective zone. Given $F_{\rm v}(\tau)=-F_{\odot}e^{-\alpha\tau}$, the atmosphere is convectively unstable if both of the following conditions are met (Appendix \ref{app3}):
\begin{equation}
\begin{split}
&\frac{\alpha}{D}<1-\beta e^{1-\beta}, \\
&\tau_{sc}<\tau_{s}, \label{atmsc}
\end{split}
\end{equation} 
where $\tau_{s}$ is the surface optical depth and $\tau_{sc}$ is the optical depth of the RCB under the Schwarzschild criterion, which is given by:
\begin{equation}
e^{\alpha\tau_{sc}}\left(1+\frac{\alpha\tau_{sc}}{\beta}\right)^{-1}=1-\frac{\alpha}{D}. \label{scrcb}
\end{equation}

However, estimating the RCB solely based on the Schwarzschild criterion might be biased because comparing the radiative-equilibrium temperature profile and the adiabat can only provide a sufficient but not necessary criterion for convection. For example, it does not apply to near-surface convective instability. The Schwarzschild criterion measures the convective instability from a top-down approach but does not consider the surface conditions such as the surface optical depth $\tau_{\rm s}$ or surface temperature $T_{\rm s}$. On terrestrial planets with surfaces, there is a temperature discontinuity between the surface and the overlying air in immediate contact with it if the system is in radiative equilibrium (see Appendix \ref{app2}). Near-surface convection will occur if the surface is warmer than the overlying atmosphere (see Equation \ref{surft}):
\begin{equation}
(2-D)S(\tau_{\rm s})+2(D-\alpha)e^{-\alpha\tau_{\rm s}}>0.
\end{equation}
For example, near-surface convection occurs using the hemi-isotropic approximation ($D=2$) and $\alpha/D<1$. This criterion differs from the Schwarzschild criterion in Equation (\ref{atmsc}) because it does not require $\tau_{sc}<\tau_{s}$. An optically thin system with a small $\tau_{\rm s}$ can still develop a deep convective zone even though it appears stable against convection under the Schwarzschild criterion. The criterion in Equation (\ref{sch}) is not applicable for estimating the RCB in this case. 

Moreover,  \cite{robinsonAnalyticRadiativeConvectiveModel2012} argued that using the Schwarzschild criterion would also lead to discontinuity of the upwelling energy fluxes at the RCB. In this study, we used a flux criterion (see below) instead of the Schwarzschild criterion to estimate the RCB. The Schwarzschild criterion was only used to ensure the radiative zone is stable against convection.

\subsection{Radiative-Convective Equilibrium Solutions}

Following \cite{robinsonAnalyticRadiativeConvectiveModel2012}, we used temperature and energy flux criteria to estimate the location of RCB and link the temperature structure of the radiative zone ($T_{\rm rad}$) with the underlying adiabat ($T_{\rm conv}$):

(1) Temperature continuity criterion:  
\begin{equation}
T_{\rm rad}(\tau_{\rm rcb})=T_{\rm conv}(\tau_{\rm rcb}),\label{tcon}
\end{equation}

(2) Upward radiative flux continuity criterion:  
\begin{equation}
F^+_{\rm rad}(\tau_{\rm rcb})=F^+_{\rm conv}(\tau_{\rm rcb}),\label{c3}
\end{equation}
where $F^+_{\rm rad}(\tau_{\rm rcb})$ and $F^+_{\rm conv}(\tau_{\rm rcb})$ are the upward \textit{radiative} fluxes at the RCB calculated using the radiative-equilibrium temperature (Equation \ref{trad}) and deep adiabatic temperature (Equation \ref{adt}), respectively.

Criterion (1) is straightforward. Criterion (2) requires that the upward radiative flux is continuous at the RCB if we neglect the convective overshoot and assume there is no vertical kinetic energy flux at the RCB. Because of the radiative flux balance $F^+_{\rm rad}(\tau_{\rm rcb})=-F_{\rm v}(\tau_{\rm rcb})-F^-_{\rm rad}(\tau_{\rm rcb})+F_{\rm int}$ in the radiative zone, Criterion (2) is equivalent to $F^+_{\rm conv}(\tau_{\rm rcb})=-F_{\rm v}(\tau_{\rm rcb})-F^-_{\rm rad}(\tau_{\rm rcb})+F_{\rm int}$ in the convective zone. It implies the energy conservation in the convective zone below the RCB. Using Criterion (2) leads to the actual RCB being located at a higher altitude than that determined by the Schwarzschild criterion (see Figure 1 in \citealt{robinsonAnalyticRadiativeConvectiveModel2012}). Previous studies have also used numerical convective adjustment schemes to ensure that the flux is continuous and the total enthalpy in the unstable region is conserved during the adjustment process (e.g., \citealt{manabeThermalEquilibriumAtmosphere1964}).

Multiple solutions for the location of the RCB may exist in some situations, but some are unphysical (\citealt{robinsonAnalyticRadiativeConvectiveModel2012}). We also require that the radiative zone is stable against convection under the Schwarzschild criterion to ensure a physically realistic solution. In Appendix \ref{app2} we derived a general form of the upward radiative flux $F^+_{\rm rad}(\tau)$ for a radiative-equilibrium atmosphere using the temperature structure (Equation \ref{trad}):
 \begin{equation}
    F_{\rm rad}^+(\tau)=\frac{1}{2}\left[\frac{S(\tau)}{D}-\frac{1}{D}\frac{\partial F_{\rm v}}{\partial\tau}-F_{\rm v}(\tau)+(2+D\tau)F_{\rm int}\right].
\end{equation}
Likewise, the upward radiative flux in the convective zone can be obtained via the integration of the deep adiabat (Equation \ref{fluxsol}). For terrestrial planets, we have:
\begin{equation}
\begin{split}
F_{\rm conv}^+(\tau)=&\frac{T_{\rm s}^4e^{D(\tau-\tau_{\rm s})}}{4}+\frac{1}{2}\int_\tau^{\tau_{\rm s}} T_{\rm s}^4\big(\frac{t}{\tau_{\rm s}}\big)^\beta e^{-D(t-\tau)}dt \\
=&\frac{T_{\rm s}^4e^{D(\tau-\tau_{\rm s})}}{4}+\\
&\frac{T_{\rm s}^4\tau_{\rm s}^{-\beta}e^{D\tau}}{2D^{\beta+1}}\big[\Gamma(\beta+1, D\tau)-\Gamma(\beta+1, D\tau_{\rm s})\big]
\end{split}
\end{equation}
where $\Gamma$ is the upper incomplete gamma function $\Gamma(s, x)=\int_x^\infty t^{s-1} e^{-t}dt$. 

For giant planets, there is no surface. The upward radiative flux in the deep convective zone is simpler:
\begin{equation}
\begin{split}
F_{\rm conv}^+(\tau)=&\frac{1}{2}\int_\tau^{\infty} T_{\rm s}^4\big(\frac{t}{\tau_{\rm s}}\big)^\beta e^{-D(t-\tau)}dt \\
=&\frac{T_{\rm s}^4\tau_{\rm s}^{-\beta}e^{D\tau}}{2D^{\beta+1}}\Gamma(\beta+1, D\tau).
\end{split}
\end{equation}

We underscore the importance of pressure as the fundamental coordinate in our atmospheric model, due to our focus on inhomogeneities across isobars. While we utilize the IR optical depth as a practical coordinate to calculate each column via $\tau=\kappa p^n$, the underlying inhomogeneity effect of opacity is investigated with a fixed pressure grid and variable $\kappa$. Consequently, $\tau_s=\kappa p_{\rm s}^n$ also adjusts since $p_{\rm s}$ remains constant.

For simplicity, we assume a direct beam ($\mu=1$) following Beer's law $F_{\rm v}(\tau)=-F_{\odot}e^{-\alpha\tau}$ where $F_{\odot}$ is the normalized local incident stellar flux. The solution with any other $\mu$ can be obtained by scaling $F_{\odot}$ and $\alpha$. We assumed $D=2$ in our calculation. As mentioned in Section \ref{sec:rad}, the system's behavior should be independent of the choice of $D$.

To explore the inhomogeneity effect, we calculate the surface temperature $T_s$ for terrestrial planets in Section \ref{tesec}. If we neglect the interior heat flow from sources such as radioactive decay or potential tidal heating, the internal heat flux stands at zero. Considering the adiabatic lapse rate defined by $\beta$, the surface temperature from the RCE solution can be expressed as $T_s (F_{\odot}, \kappa, \alpha)$, which is a function of the incident stellar flux $F_{\odot}$, the IR opacity coefficient $\kappa$, and the visible-to-IR opacity ratio $\alpha$. The surface IR optical depth $\tau_s$ alters with changes in $\kappa$. We will scrutinize the effect of each parameter within the set $\left\{ F_{\odot}, \kappa, \alpha \right\}$ individually while maintaining the other parameters constant in the RCE column model.

In the context of giant planets, all columns share a common adiabat in the deep interior. Given this deep adiabat, with $T_{\rm s}$ at $p_{\rm s}$, we can compute the internal heat flux $F_{\rm int} (F_{\odot}, \kappa, \alpha)$ at the top of the atmosphere to probe the inhomogeneity effect in Section \ref{gsec}.

\section{Terrestrial planets} \label{tesec}

The inhomogeneity in a planet's atmosphere affects its average surface temperature, an important factor in determining its potential habitability. We will only focus on ``thick" atmospheres, which have convection zones above the surface. ``Thin" atmospheres in radiative equilibrium, which have a temperature jump near the cold surface, thermal conduction would occur (see Appendix \ref{app2}). We will not consider this scenario as the surface temperature may be primarily influenced by the surface rather than the atmosphere.

Using the RCE model, we explored the response of the surface temperature $T_s$ to changes in each parameter within the set $\left\{ F_{\odot}, \kappa, \alpha \right\}$. Firstly, we maintained constant opacity and adjusted the incident stellar flux $F_{\odot}$(\Cref{fig:Teresponse}A); secondly, we modified the $\kappa$ while keeping the incident stellar flux and the visible-to-IR opacity ratio constant (\Cref{fig:Teresponse}B); thirdly, we maintained the incident stellar flux and IR opacity constant but altered the visible opacity by adjusting the visible-to-IR opacity ratio $\alpha$ (\Cref{fig:Teresponse}C); finally, we fixed constant incident stellar flux and visible opacity, while varying the IR opacity $\kappa$ (\Cref{fig:Teresponse}D). Subsequently, we demonstrated examples of how changes in incident stellar flux and/or atmospheric opacity can affect surface temperature by contrasting two non-uniform atmospheric columns to a uniform column (\Cref{fig:tscaling} and \Cref{tecontour}).

\subsection{Response Functions of Surface Temperature}

\begin{figure*}
  \centering \includegraphics[width=0.85\textwidth]{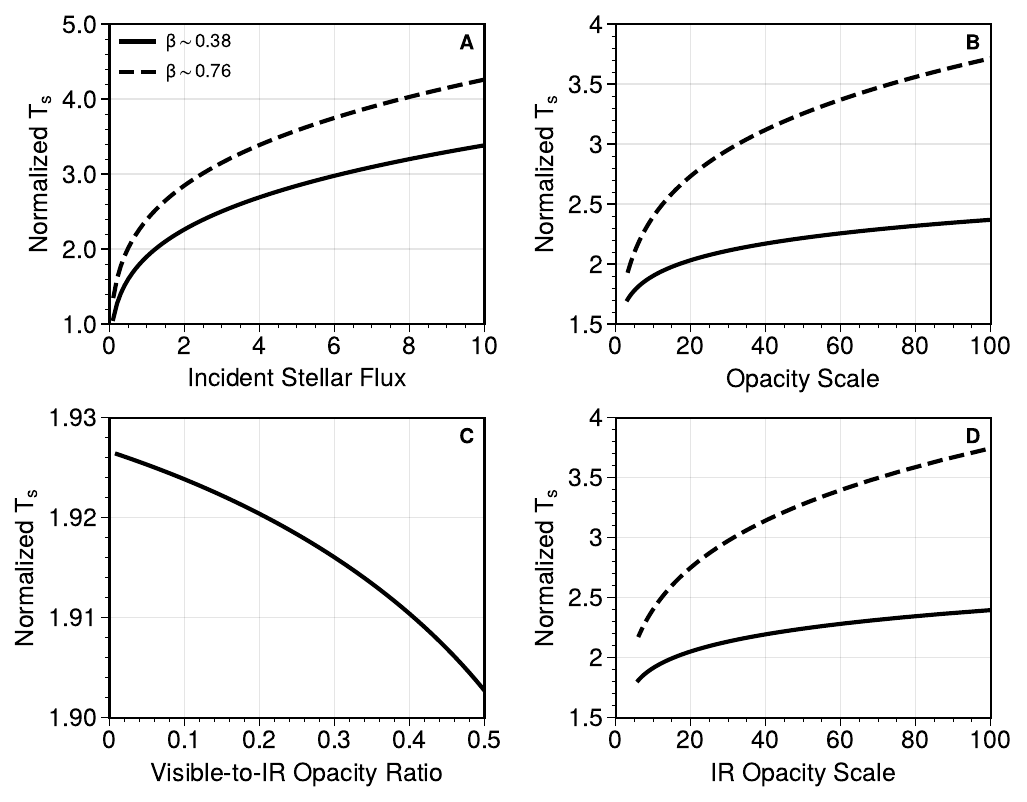} 
  \caption{Concave response functions of surface temperature in relation to the incident stellar flux and opacity on terrestrial planets, with the diffusivity parameter $D$ set to 2. In Panels A and B, the solid lines represent the visible-to-IR opacity ratio $\alpha=0.5$ and the adiabatic slope parameter $\beta\sim0.38$. The dashed lines, on the other hand, correspond to $\alpha=0.05$ and $\beta\sim0.76$. Panel A illustrates how the surface temperature changes with the incident stellar flux when the opacity is kept constant, corresponding to a surface IR optical depth of $\tau_{\rm s}=3$. Panel B demonstrates the variation in surface temperature as the IR opacity coefficient $\kappa$ increases, with the surface IR optical depth $\tau_{\rm s}$ changing from 3 to 100. In this panel, the incident stellar flux ($F_{\odot}=1$) is held constant, and we also kept $\alpha$ fixed, ensuring that the visible and IR opacity increase by the same factor. Panel C exhibits the alteration in the surface temperature as the visible-to-IR opacity ratio increases. In this case, we maintain the incident stellar flux ($F_{\odot}=1$) and $\kappa$ (corresponding to $\tau_{\rm s}=10$) as fixed. Here, only one scenario ($\beta\sim0.38$) of varying $\alpha$ is displayed because the maximum $\alpha$ (approximately 0.0665) in the other case is too small to be plotted on the same graph. Panel D mirrors Panel B, with fixed $F_{\odot}=1$ and varying IR opacity coefficient $\kappa$. However, in this panel, we also vary $\alpha$ so that the visible opacity ($\alpha\tau$) remains unaffected by the IR opacity ($\tau$). We ensure the product $\alpha \kappa$ is constant, implying $\alpha\tau_{\rm s}=3$ for the $\beta\sim0.38$ case (solid) and $\alpha\tau_{\rm s}=0.3$ for $\beta\sim0.76$ case (dashed).} \label{fig:Teresponse}
\end{figure*}

For terrestrial planets with $F_{\rm int}=0$, the temperature continuity constraint of the RCB (Equation \ref{tcon}) becomes:
 \begin{equation}
   S(\tau_{\rm rcb})=T_{\rm s}^4\left(\frac{\tau_{\rm rcb}}{\tau_{\rm s}}\right)^\beta,\label{tetcon}
\end{equation}

The flux continuity constraint $F^+_{\rm rad}(\tau_{\rm rcb})=F^+_{\rm conv}(\tau_{\rm rcb})$ (Equation \ref{c3}) becomes:

\begin{equation}
\begin{split}
&S(\tau_{\rm rcb})-\frac{\partial F_{\rm v}}{\partial\tau}\Big|_{\tau_{\rm rcb}}-DF_{\rm v}(\tau_{\rm rcb})
=T_{\rm s}^4e^{D(\tau_{\rm rcb}-\tau_{\rm s})}\\
&\left\{\frac{D}{2}+(D\tau_{\rm s})^{-\beta}e^{D\tau_{\rm s}}
\left[\Gamma(\beta+1, D\tau_{\rm rcb})-\Gamma(\beta+1, D\tau_{\rm s})\right]\right\}
\end{split}
\end{equation}

We can further eliminate $T_{\rm s}^4$ to obtain the $\tau_{\rm s}-\tau_{\rm rcb}$ relation:
\begin{equation}
\begin{split}
1&-\frac{1}{S(\tau_{\rm rcb})}\frac{\partial F_{\rm v}}{\partial\tau}\Big|_{\tau_{\rm rcb}}-\frac{DF_{\rm v}(\tau_{\rm rcb})}{S(\tau_{\rm rcb})}-\frac{D\tau_{\rm s}^{\beta}e^{D(\tau_{\rm rcb}-\tau_{\rm s})}}{2\tau_{\rm rcb}^\beta} \\
&=\frac{e^{D\tau_{\rm rcb}}}{\tau_{\rm rcb}^\beta D^{\beta}}
\left[\Gamma(\beta+1, D\tau_{\rm rcb})-\Gamma(\beta+1, D\tau_{\rm s})\right].\label{ts-trcb}
\end{split}
\end{equation}
Given $D=2$ and the adiabatic slope parameter $\beta$, we chose appropriate values of $\alpha$ and $\kappa$ (thus $\tau_{\rm s}$) that satisfy the convectively unstable requirement in \Cref{atmsc} and solved for $\tau_{\rm rcb}$ and $T_{\rm s}$ using root-finding algorithms. 

Two different values of $\beta$ were used: $\beta\sim0.76$ representing a dry atmosphere with diatomic gas and $n=1.5$, and $\beta\sim0.38$ representing an atmosphere composed of more complex molecules and/or with a moist-adiabatic convective zone. According to \Cref{atmsc}, a smaller $\beta$ allows for a larger $\alpha$.

The dependence of the surface temperature on the incident stellar flux is straightforward. Because $F_{\rm v}$ is an exponential function, its derivatives, integrals, and the stellar heating source terms $S$ in \Cref{srceq} are proportional to $F_{\odot}$. As seen from \Cref{ts-trcb}, given $\alpha$ and $\tau_{\rm s}$, $\tau_{\rm rcb}$ can be determined and the value is independent of $F_{\odot}$. According to \Cref{tetcon}, the surface temperature response function can be written as:
 \begin{equation}
   T_{\rm s}(F_{\odot})\propto S(\tau_{\rm rcb})^{1/4} \propto F_{\odot}^{1/4}, \label{TeonI}
\end{equation}
The response functions in the form of $y=x^{1/4}$ are concave in shape. The temperature increases with the incident stellar flux when the opacity is held constant (\Cref{fig:Teresponse}A). $T_{\rm s}$ with a larger $\beta$ shows a higher surface temperature due to the steeper atmospheric adiabat. 

We now examine the response of the surface temperature $T_{\rm s}$ to the opacity when the incident stellar flux is held as $F_{\odot}=1$. \Cref{fig:Teresponse}B shows that $T_{\rm s}$ increases with the IR opacity coefficient $\kappa$ while keeping the $\alpha$ the same (i.e., both visible and IR opacity increase). If $\beta\sim0.76$ and $\alpha=0.05$, the increase can be more than a factor of 2 as $\kappa$ increases by 30, corresponding to the surface IR optical depth $\tau_{\rm s}$ changes from 3 to 100. The increase is smaller if $\beta\sim0.38$ and $\alpha=0.5$. This is the greenhouse effect caused by increased IR opacity. 

But the temperature increase in the RCE framework is different from the traditional description of the greenhouse effect in radiative-equilibrium calculations because the lapse rate is different in RE versus RCE (\citealt{pierrehumbertPrinciplesPlanetaryClimate2010}, p.211). When opacity is not large, as the IR opacity increases, temperature increases, and more IR fluxes are emitted from below. The RCB moves down (i.e., $\tau_{\rm rcb}$ increases) to radiate more flux to space. When opacity further increases, both the IR flux and $\tau_{\rm rcb}$ are not sensitive to the bottom part of the convective zone with a large opacity. In the optically thick limit as $\tau_{\rm s}\rightarrow \infty$, $\tau_{\rm rcb}$ reaches an asymptotic value, which is given by:
\begin{equation}
\frac{e^{D\tau_{\rm rcb}}}{\tau_{\rm rcb}^\beta D^{\beta}}
\Gamma(\beta+1, D\tau_{\rm rcb})+\frac{1}{S(\tau_{\rm rcb})}\frac{\partial F_{\rm v}}{\partial\tau}\Big|_{\tau_{\rm rcb}}+\frac{DF_{\rm v}(\tau_{\rm rcb})}{S(\tau_{\rm rcb})}=1.
\end{equation}
The asymptotic value of $\tau_{\rm rcb}$ is $\sim 0.4259$ for $\beta\sim 0.38$ and $\alpha=0.5$. $\tau_{\rm rcb}\sim 1.8908$ for $\beta\sim 0.76$ and $\alpha=0.05$. Using \Cref{tetcon}, we thus infer that, for high values of $\kappa$ (thus high value of $\tau_{\rm s}$), the surface temperature scales as:
\begin{equation}
T_{\rm s}\propto \tau_{\rm s}^{\beta/4}\propto \kappa^{\beta/4}.
\end{equation}
Because $\beta/4=\zeta(\gamma-1)/n\gamma<1$, $T_{\rm s}(\kappa)$ is a concave function in the limit of $\tau_{\rm s}\rightarrow \infty$. \Cref{fig:Teresponse}B shows that $T_{\rm s}(\kappa)$ is a concave function for both small and large $\kappa$.

We also examined the dependence of the surface temperature on the visible opacity only. We kept the $\kappa$ the same and varied $\alpha$. The surface temperature decreases as $\alpha$ increases (\Cref{fig:Teresponse}C). This is the anti-greenhouse effect (e.g., \citealt{mckayThermalStructureTitan1989}), which occurs when less stellar flux reaches the lower atmosphere and the surface due to increased absorption in the upper atmosphere as visible opacity increases. The surface temperature is concave but very weakly dependent on $\alpha$.

\begin{figure*}
  \centering \includegraphics[width=0.85\textwidth]{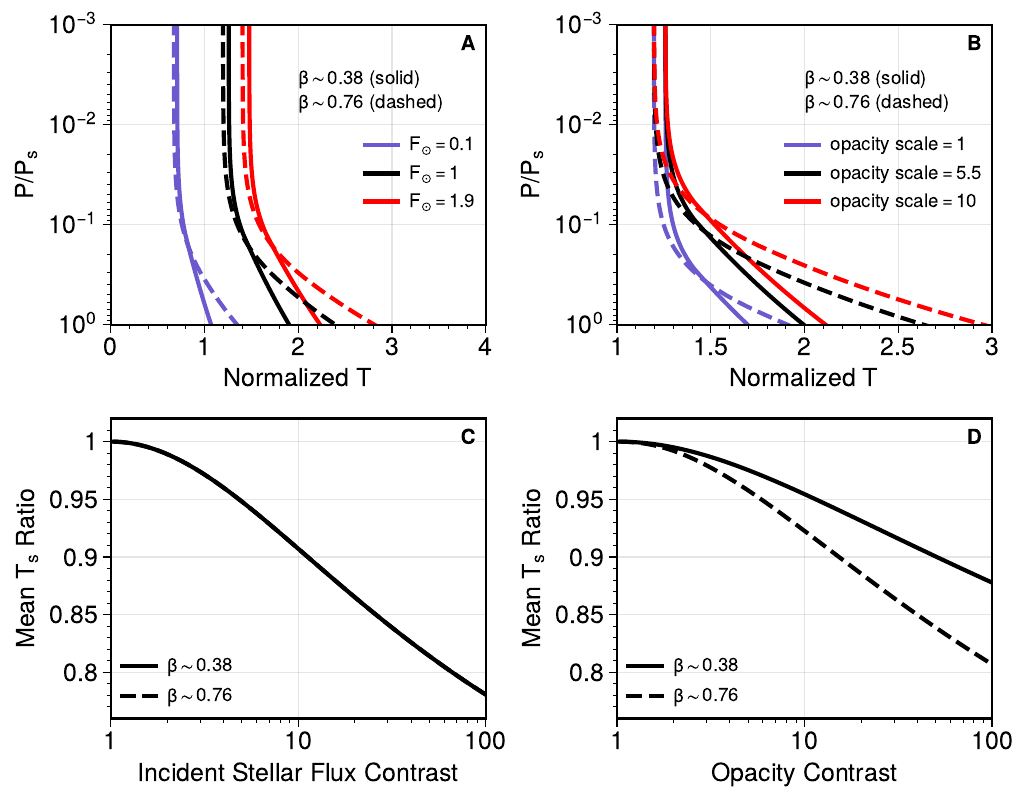} 
  \caption{Mixing of two inhomogeneous RCE columns for terrestrial cases, with the diffusivity parameter $D$ set at 2. Similar to \Cref{fig:Teresponse}, the solid lines correspond to the visible-to-IR opacity ratio $\alpha=0.5$ and the adiabatic slope parameter $\beta\sim0.38$, while the dashed lines are for $\alpha=0.05$ and $\beta\sim0.76$. The left panels show scenarios with inhomogeneous incident stellar flux, with the IR opacity coefficient $\kappa$ maintained constant, corresponding to the surface IR optical depth $\tau_{\rm s}=3$. The right panels present cases with inhomogeneous opacities, wherein the same scaling factor is applied to both visible and IR opacity while keeping the incident stellar flux constant at $F_{\odot}=1$. Panel A illustrates the temperature profiles as a function of normalized pressure ($p/p_{\rm s}$) for normalized incident stellar flux $F_{\odot}=0.1$ (blue), 1 (black, the mixed column), and 1.9 (red). Panel B shows the temperature profiles as a function of normalized pressure when scaling the IR opacity coefficient $\kappa$ by a factor of 1 (blue), 5.5 (black, the mixed column), and 10 (red). This corresponds to the surface IR optical depth $\tau_{\rm s}$ of 3, 16.5, and 30, respectively. Panel C displays the ratios of the averaged surface temperature $T_{\rm s}$ between the inhomogeneous columns and the homogeneous column as the contrast of incident stellar flux increases, with $\kappa$ held constant. The solid and dashed curves overlap in this panel. Panel D presents the ratios of the averaged surface temperature $T_{\rm s}$ between the inhomogeneous and the homogeneous columns as the opacity contrast amplifies, with a fixed incident stellar flux.} \label{fig:tscaling}
\end{figure*}

\begin{figure}
  \centering \includegraphics[width=0.45\textwidth]{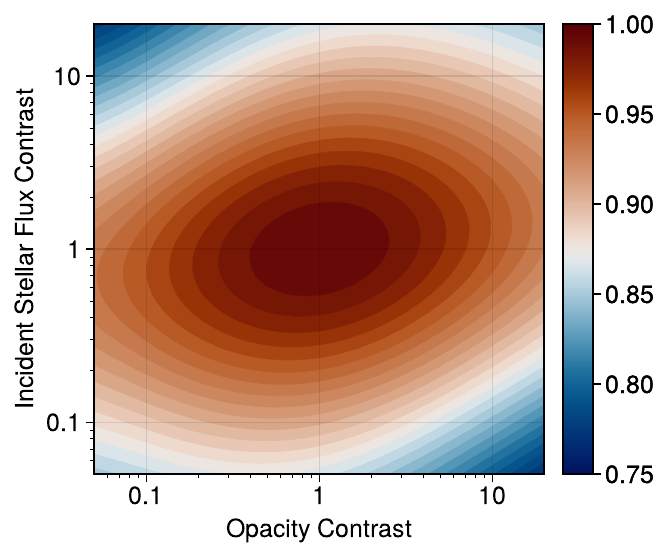} 
  \caption{The ratio of the averaged surface temperature $T_{\rm s}$ in the inhomogeneous columns compared to the homogeneous column as a function of the incident stellar flux and opacity contrast between the two columns. We adopted the diffusivity parameter $D=2$, visible-to-IR opacity ratio $\alpha=0.5$, and the adiabatic slope parameter $\beta\sim 0.38$. The reference incident stellar flux is set at $F_{\odot}=1$, and the incident stellar flux of the second column varies from 0.05 to 20. The reference IR opacity coefficient $\kappa$ is set so that the surface IR optical depth $\tau_{\rm s}$ is 20, and $\tau_{\rm s}$ in the second column varies from 1 to 400.} \label{tecontour}
\end{figure}

Likewise, we test cases with IR opacity only by increasing $\kappa$ and decreasing $\alpha$ but keeping their product ($\alpha\kappa$) the same (\Cref{fig:Teresponse}D). The surface temperature is only slightly higher than the case with $\alpha$ fixed (\Cref{fig:Teresponse}B). The response function, again, is concave. 

According to Jensen's inequality, the concave nature of the response functions of the surface temperature implies that the inhomogeneity of incident stellar flux and/or atmospheric opacity would lower the average surface temperature on terrestrial planets. This is the terrestrial planet version of the inhomogeneity effect.

\subsection{Surface Temperature Change with Increasing Inhomogeneity}

To illustrate the contrast between inhomogeneous and homogeneous atmospheric conditions, we can compare the mean surface temperature of two columns with different incident stellar flux and/or opacity to a homogeneous case in which the incident stellar flux and opacity are the averages of the two columns at a constant pressure level. By varying the degree of inhomogeneity between the two columns, we can explore how the mean surface temperature changes with the inhomogeneity. This simple example can also be extrapolated to understand the behavior of multiple atmospheric columns.

We designed two sets of comparison cases: $\{\alpha=0.5,\beta\sim0.38\}$ and $\{\alpha=0.05, \beta\sim0.76\}$, respectively. For each $\beta$, the incident stellar flux values for the two columns are 0.1 and 1.9, respectively, and the average incident stellar flux for the mixed column is 1. In all cases, the IR opacity coefficient $\kappa$ is held constant so that the surface IR optical depth $\tau_{\rm s}=3$. The resulting temperature profiles are shown in \Cref{fig:tscaling}A. As aforementioned, our model essentially uses pressure as the coordinates, so we presented the results using the normalized pressure $p/p_{\rm s}$. As the incident stellar flux decreases, the atmosphere becomes colder. The cases with a larger $\beta$ value exhibit steeper adiabats in the lower convective atmosphere.

Then we selected a reference column with an incident stellar flux value of $F_{\odot}=1$ and surface IR optical depth $\tau_{\rm s}=3$. We then gradually increased the contrast between the two columns by varying the value in the second column from 1 to 100. Because the response functions of the surface temperature are concave, the ratio of the averaged $T_{\rm s}$ between the inhomogeneous columns to that in the uniform column is smaller than unity (\Cref{fig:tscaling}C). As the incident stellar flux contrast increases, the ratio between the inhomogeneous columns and the uniform column decreases. With the adopted parameters, if the incident stellar flux contrast exceeds 60-fold, the mean surface temperature in the inhomogeneous case can be more than 20\% lower than the uniform column. The trend of the surface temperature ratio is the same regardless of the value of $\beta$, as the surface temperature dependence on incident stellar flux does not depend on $\beta$ (Equation \ref{TeonI}).

The decreasing trend of the $T_{\rm s}$ ratio with increasing column contrast (\Cref{fig:tscaling}C) cannot be explained by Jensen's inequality, which only compares the inhomogeneous case with the homogeneous case. A more advanced theory is needed to compare two inhomogeneous cases, which will be covered in the next paper (\citealt{zhangInhomogeneityEffectII2023}). However, for the simplest case of only two data points (columns), it can be shown that the mean value of a convex (concave) function increases (decreases) as the degree of inhomogeneity increases. The mathematical proof is provided in Appendix \ref{app:convex}.

Tests of opacity inhomogeneity are shown in \Cref{fig:tscaling}B, where the incident stellar flux $F_{\odot}$ was fixed at 1, and the IR opacity scale coefficient $\kappa$ is scaled so that $\tau_{\rm s}$ was set to 3, 16.5, and 30 for the three cases, respectively. The visible-to-IR opacity ratio $\alpha$ was also fixed so that the visible opacity changes by the same factor. In the normalized pressure coordinate, the atmosphere becomes hotter as the opacity increases. Then we used $\tau_{\rm s}=3$ case as a reference and varied $\kappa$ by a factor of 1 to 100 in the second column. Similar to the incident stellar flux test, the average $T_{\rm s}$ between the inhomogeneous columns is lower than the uniform column (\Cref{fig:tscaling}D). As the opacity contrast increases, the $T_{\rm s}$ ratio of the inhomogeneous to the homogenous case decreases. Depending on $\beta$, the mean $T_{\rm s}$ in the inhomogeneous case is 10-20\% lower than the uniform column when the opacity contrast reaches 10-to-100 fold.

We also tested various combinations of incident stellar flux and opacity contrast between the two columns. The reference column was chosen with $D=2$, $\alpha=0.5$, $\beta\sim0.38$, $F_{\odot}=1$, and $\kappa$ corresponding to $\tau_{\rm s}=20$. Then the values of $F_{\odot}$ and $\kappa$ were varied (while keeping $\alpha$ the same) in the second column by up to 20 to cover a wide range of parameters. Note that the contrast would increase as the value deviates from unity, whether smaller or larger. The ratio of the average $T_{\rm s}$ in the two columns to the uniform column is shown in \Cref{tecontour}. The ratio is always less than unity, confirming that the inhomogeneous stellar flux and/or opacity will always lead to more surface cooling. 

The pattern of the ratio appears centrosymmetric about the unity-contrast point. As expected, if one contrast is fixed and the other is varied, the surface temperature decreases with increasing contrast. However, the combination of incident stellar flux and opacity contrast can either enhance or compensate for each other's effects. If one column has a higher incident stellar flux and a lower opacity than the other, the temperature decrease would be even larger. For example, the ratio reaches above 25\% for extremely inhomogeneous columns if we increase $\kappa$ by 20 and decrease $F_{\odot}$ by 20. However, if one column has a higher stellar flux as well as a larger opacity, the temperature decrease is smaller (15\%, \Cref{tecontour}). The detailed mechanism is yet to be explored as it is difficult to compare two inhomogeneous cases with two independent variables ($F_{\odot}$ and $\kappa$). 

The two-column experiment can be extrapolated to understand the scenario of multiple columns. According to Jensen's equality, averaging multiple inhomogeneous columns should also lead to the mean-temperature reduction. The decrease in the mean temperature would depend on the distribution of the inhomogeneity in the atmosphere and the surface. For reference, comprehensive 3D simulations of synchronously rotating planets in \cite{leconte3DClimateModeling2013} also showed that the mean surface temperature of the 3D models is about 10-20\% lower than the 1D model results (see their Figure 2).

\section{Giant planets} \label{gsec}

Giant planets do not have surfaces. We assume that the atmospheric temperature is horizontally uniform in the deep convective layer. This assumption is based on the idea that convection would strongly homogenize the horizontal temperature distribution in the interior over an isobar (i.e., constant pressure level). Using the RCE column model, we can solve for $F_{\rm int}(F_{\odot}, \kappa, \alpha)$ and investigate the effect of atmospheric inhomogeneity of $\left\{ F_{\odot}, \kappa, \alpha \right\}$ on the outgoing heat flux at the top of the atmosphere. 

\begin{figure*}
\centering \includegraphics[width=0.85\textwidth]{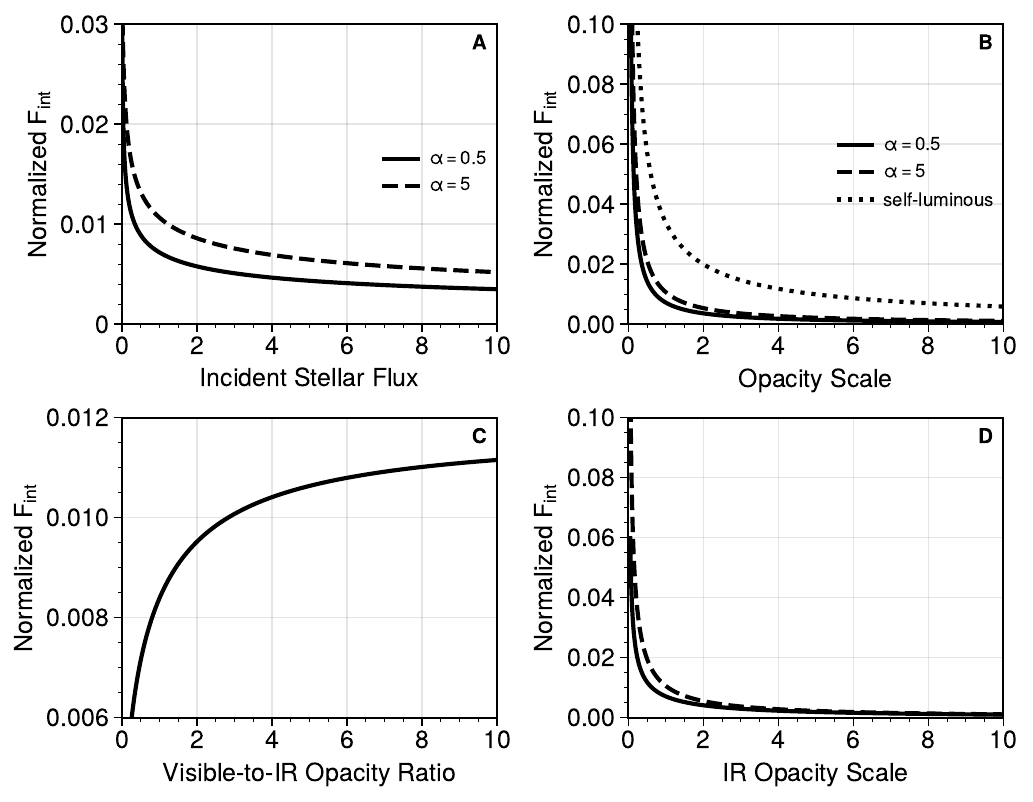}
\caption{Response functions of outgoing internal heat flux $F_{\rm int}$ to incident stellar flux and opacity on giant planets. We selected the diffusivity parameter $D=2$ and the adiabatic slope parameter $\beta\sim 0.76$. Two irradiated planet cases are displayed with the visible-to-IR opacity ratio $\alpha=0.5$ (solid) and $\alpha=5$ (dashed). An additional self-illuminated planet case is shown in Panel B. Panel A represents the variation of the internal heat flux with the incident stellar flux $F_{\odot}$. Here, the opacity is kept constant with a fixed $K=0.2$ (refer to Equation \ref{Kdef} for the definition of $K$). Panel B demonstrates the variation of the internal heat flux with the IR opacity coefficient $\kappa$, with the incident stellar flux held constant at $F_{\odot}=1$. As $\alpha$ is fixed, the visible opacity varies by the same factor for irradiated planets (solid and dashed). To retain the same interior temperature at the same pressure, $K$ is scaled with $\kappa$ via $K\propto\kappa^{-\beta}$ for all planets. Panel C presents the change in the surface temperature as the visible-to-IR opacity ratio escalates, with fixed incident stellar flux ($F_{\odot}=1$) and $K=0.2$. Panel D mirrors Panel B, with a fixed $F_{\odot}=1$ and varying IR opacity coefficient $\kappa$ (and $K$ accordingly), but here $\alpha$ is also varied such that the visible opacity ($\alpha\tau$) does not alter with the IR opacity ($\tau$). The product $\alpha \kappa$ is held constant for both cases. The response functions are convex in Panels A, B, and D and concave in Panel C.} \label{fig:Firesponse}
\end{figure*}

\begin{figure*}
\centering \includegraphics[width=0.85\textwidth]{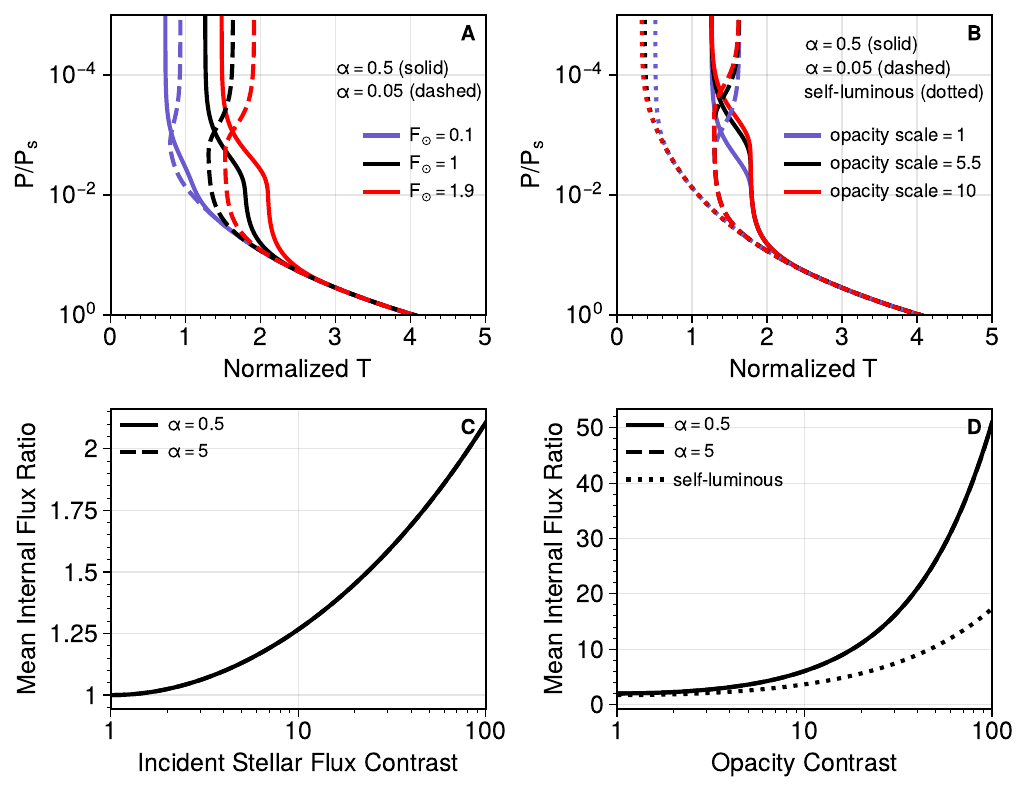}
\caption{Mixing of two inhomogeneous RCE columns for giant planet cases. The diffusivity parameter is chosen as $D=2$, and the adiabatic slope parameter is $\beta\sim 0.76$. Two cases of irradiated planets are presented with visible-to-IR opacity ratios $\alpha=0.5$ (solid) and $\alpha=5$ (dashed), alongside an additional case of a self-illuminated planet shown in Panel B. The left panels display cases with inhomogeneous incident stellar flux while maintaining constant opacity and $K=0.2$. The right panels show cases with inhomogeneous opacities for irradiated (solid and dashed) and self-luminous bodies (dotted), with constant incident stellar flux set as $F_{\odot}=1$. Panel A: Temperature profiles plotted as a function of normalized pressure ($p/p_{\rm s}$) for normalized incident stellar flux $F_{\odot}=0.1$ (blue), 1 (black, the mixed column), and 1.9 (red). Panel B: Temperature profiles as a function of normalized pressure for IR opacity coefficient $\kappa$ scaled by factors of 1 (blue), 5.5 (black, the mixed column), and 10 (red) with corresponding adjustments to $K$. Panels C and D present the ratios of the average $F_{\rm int}$ in the inhomogeneous columns compared to the homogeneous column when varying the contrast of the incident stellar flux (with fixed opacity) and opacity (with fixed stellar flux), respectively. The dashed and dotted curves overlap in Panel C.} \label{fscaling}
\end{figure*}

\begin{figure}
\centering \includegraphics[width=0.45\textwidth]{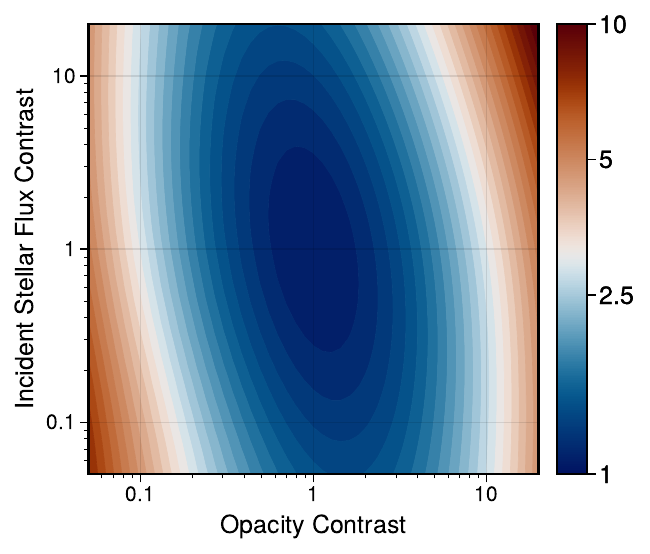}
\caption{The ratio of the averaged outgoing internal heat flux in the inhomogeneous columns compared to the homogeneous column as a function of the incident stellar flux and opacity contrast between the two columns. We have selected the diffusivity parameter $D=2$, visible-to-IR opacity ratio $\alpha=1$, and the adiabatic slope parameter $\beta\sim 0.76$. The reference incident stellar flux is set at $F_{\odot}=1$, and the reference IR opacity coefficient $\kappa$ is 1, corresponding to $K=0.2$. In the second column, both quantities were varied by a factor ranging from 0.05 to 20.} \label{fcontour}
\end{figure}

Similar to the terrestrial scenarios, we investigate four different situations, each with a specific deep adiabat. Firstly, we keep the opacity constant and vary the incident stellar flux $F_{\odot}$ (\Cref{fig:Firesponse}A); secondly, we alter $\kappa$ while maintaining constant incident stellar flux and visible-to-IR opacity ratio (\Cref{fig:Firesponse}B); thirdly, we keep the incident stellar flux and IR opacity constant but vary the visible opacity by altering the visible-to-IR opacity ratio $\alpha$ (\Cref{fig:Firesponse}C); Fourthly, we maintain constant incident stellar flux and visible opacity and adjust the IR opacity $\kappa$ (\Cref{fig:Firesponse}D). Lastly, we illustrate the impacts of changes in incident stellar flux and/or atmospheric opacity on the outgoing internal heat flux by contrasting two non-uniform atmospheric columns to a uniform one (\Cref{fscaling} and \Cref{fcontour}). As per our theory, the observed internal heat flux in an inhomogeneous atmosphere would exceed that in a uniform one.

\subsection{Response Functions of Outgoing Internal Flux}

The interior adiabat is characterized by $T_{\rm s}$ at a deep reference pressure $p_{\rm s}$ (or $\tau_{\rm s}$). Because the choice of the reference pressure and temperature is somewhat arbitrary on giant planets, here we introduce a new variable to simplify our discussion:
\begin{equation}
K=T_{\rm s}^4\tau_{\rm s}^{-\beta}.
\label{Kdef}
\end{equation}
$K$ is a proxy that measures the normalized interior temperature (or entropy) in the optical depth coordinates such that \Cref{adt} becomes $T_{\rm conv}^4(\tau)=K\tau^\beta$. One can consider $K^{1/4}$ as the temperature at the $\tau=1$ level following the deep adiabat. Because $\tau=\kappa p^n$, assuming homogeneous interior temperature $T_{\rm s}$ over a constant pressure surface implies that $K\kappa^\beta$ is a constant. Thus when $\kappa$ changes over the isobar, $K$ should scale as $\kappa^{-\beta}$ accordingly. If $\kappa$ is taken as a constant, $K$ is a constant in the interior.

The temperature continuity constraint of RCB (Equation \ref{tcon}) is:
 \begin{equation}
   S(\tau_{\rm rcb})+(D+D^2\tau_{\rm rcb})F_{\rm int}=K\tau_{\rm rcb}^\beta. \label{fitcon}
\end{equation}
The flux continuity constraint $F^+_{\rm rad}(\tau_{\rm rcb})=F^+_{\rm conv}(\tau_{\rm rcb})$ (Equation \ref{c3}) at RCB becomes:
\begin{equation}
\begin{split}
&S(\tau_{\rm rcb})-\frac{\partial F_{\rm v}}{\partial\tau}\Big|_{\tau_{\rm rcb}}-DF_{\rm v}(\tau_{\rm rcb})+(2D+D^2\tau_{\rm rcb})F_{\rm int}\\
&=KD^{-\beta}e^{D\tau_{\rm rcb}}\Gamma(\beta+1, D\tau_{\rm rcb}). \label{fifcon}
\end{split}
\end{equation}

Like in the case of terrestrial planets, given $\beta, D, K,$ and $F_{\rm v}$, $\tau_{\rm rcb}$ and $F_{\rm int}$ can be solved using root-finding algorithms. But the giant planet case is simpler because the response functions of $F_{\rm int}$ can be explicitly obtained, as shown below. 

If there is no external heating source, $F_{\rm v}=S=0$ in \Cref{srceq}. It occurs on a self-luminous planet without incident stellar flux, like a free-floating planet or a brown dwarf, or the nightside of a tidally locked planet with zero heat redistribution from the dayside. In this situation, the optical depth at the RCB ($\tau_{\rm rcb}$) on a self-luminous body---donated as $\tau_{\rm rcb}^*$---can be directly solved by: 
\begin{equation}
(D\tau_{\rm rcb}^*)^{-\beta}e^{D\tau_{\rm rcb}^*}\Gamma(\beta+1, D\tau_{\rm rcb}^*)-(1+D\tau_{\rm rcb}^*)^{-1}=1.
\end{equation}
Given $D=2$ and $\beta\sim 0.76$ for dry hydrogen-dominated atmospheres, $\tau_{\rm rcb}^*\sim1.2178$. It is a constant value that is independent of the interior entropy (e.g., $K$). The $F_{\rm int}-K$ relation can be obtained by:
\begin{equation}
F_{\rm int}=\frac{\tau_{\rm rcb}^{*\beta}}{D+D^2\tau_{\rm rcb}^*}K\propto\kappa^{-\beta}.
\end{equation}
In the above derivation, we have used that $K\kappa^\beta$ is a constant in the deep convective zone. Given $D=2$ and $\beta\sim 0.76$, $F_{\rm int}\sim0.1691 K$. The dependence of $F_{\rm int}$ on the IR opacity is shown in \Cref{fig:Firesponse}B with $K=0.2$. It is a convex function since $\beta$ is positive. 

For irradiated planets, we can also find the $F_{\rm int}-\tau_{\rm rcb}$ and $K-\tau_{\rm rcb}$ relations from \Cref{fitcon} and \Cref{fifcon}:
\begin{equation}
\begin{split}
F_{\rm int}&=\frac{S(\tau_{\rm rcb})(D+D^2\tau_{\rm rcb})^{-1}+D^{-1}\frac{\partial F_{\rm v}}{\partial\tau}\Big|_{\tau_{\rm rcb}}+F_{\rm v}(\tau_{\rm rcb})}{2+D\tau_{\rm rcb}-(1+D\tau_{\rm rcb})\tau_{\rm rcb}^{-\beta}D^{-\beta}e^{D\tau_{\rm rcb}}\Gamma(\beta+1, D\tau_{\rm rcb})}\\
&-S(\tau_{\rm rcb})(D+D^2\tau_{\rm rcb})^{-1}, \label{fitau}
\end{split}
\end{equation}
\begin{equation}
K=\frac{S(\tau_{\rm rcb})(1+D\tau_{\rm rcb})^{-1}+\frac{\partial F_{\rm v}}{\partial\tau}\Big|_{\tau_{\rm rcb}}+DF_{\rm v}(\tau_{\rm rcb})}{\tau_{\rm rcb}^{\beta}[1+(1+D\tau_{\rm rcb})^{-1}]-D^{-\beta}e^{D\tau_{\rm rcb}}\Gamma(\beta+1, D\tau_{\rm rcb})}. \label{ktau}
\end{equation}

The response functions of $F_{\rm int}$ can be obtained without numerically solving for $\tau_{\rm rcb}$. We first consider the dependence of $F_{\rm int}$ on $F_{\odot}$ assuming the opacity is horizontally uniform. A constant $K\kappa^\beta$ implies a constant $K$. According to Equations (\ref{fitau}) and (\ref{ktau}), the ratios $F_{\rm int}/K$ and $F_{\odot}/K$ are not explicit functions of $F_{\odot}$ and only depend on $\tau_{\rm rcb}$. Given a constant $K$, by surveying $\tau_{\rm rcb}$ from $\tau_{\rm rcb}^*$\footnote{The minimum value of $\tau_{\rm rcb}$ is $\tau_{\rm rcb}^*$ in the self-luminous case because the external heating source would only push the RCB deeper into the interior on irradiated planets.} to large, we can calculate the corresponding $F_{\rm int}$ versus $F_{\odot}$. The resultant $F_{\rm int}-F_{\odot}$ relation is plotted in \Cref{fig:Firesponse}A. The internal heat flux decreases with increasing incident stellar flux, indicating a strong suppression of interior cooling by external heating. The response function is convex for both $\alpha=0.5$ and 5.

Next, we calculate the dependence of $F_{\rm int}$ on the opacity represented by $\kappa$, assuming a constant incident stellar flux of $F_{\odot}=1$. In this situation, a constant $K\kappa^\beta$ over the isobar implies $K\propto\kappa^{-\beta}$. By surveying $\tau_{\rm rcb}$ from small to large, we can calculate $F_{\rm int}$ via \Cref{fitau} and $K$ via \Cref{ktau}, and translate the $F_{\rm int}-K$ relation to the $F_{\rm int}-\kappa$ relation. The response function is shown in \Cref{fig:Firesponse}B with fixed $\alpha$. $F_{\rm int}$ decreases strongly with increasing opacity. This is because the internal temperature ($K$) decreases as $\kappa$ increases. A lower internal entropy results in a deeper RCB and less radiative cooling to space. With the same IR opacity, the emitted internal heat flux is smaller on irradiated planets than on self-luminous bodies because the irradiation pushes the RCB to a deeper atmosphere and suppresses interior cooling. The dependence of $F_{\rm int}$ on opacity is also convex.

If we fix the IR opacity but change the visible opacity, the response function must be numerically solved using root-finding algorithms. The outgoing internal heat flux $F_{\rm int}$ increases with $\alpha$ (\Cref{fig:Firesponse}C). This is because a strong absorption of the stellar flux at the upper atmosphere would move the RCB higher and increases infrared cooling. The dependence is not strong as the internal heat flux only changes by less than a factor of 2 when $\alpha$ varies by more than a factor of 20. 

Interestingly, the response functions of $F_{\rm int}$ to incident stellar flux and IR opacity are all convex, but it is a concave function of $\alpha$. One way to explain the concavity is to look at the second derivative of the attenuated incident stellar flux at a given IR optical depth $\tau$: 
\begin{equation}
\frac{\partial^2}{\partial^2 \alpha}\left(-F_{\odot}e^{-\alpha\tau}\right)=-F_{\odot}\tau^2 e^{-\alpha\tau} < 0.
\end{equation}
The downward stellar flux is a concave function of $\alpha$. The inhomogeneity of visible opacity would enhance the stellar radiation at the RCB in the homogeneous case and push it to a slightly lower altitude. Consequently, the outgoing internal flux is smaller. In other words, the inhomogeneity of visible opacity would heat the interior. However, this heating effect is small because the RCB optical depth is usually so large that the function $e^{-\alpha\tau}$ is approximately linear (i.e., the secondary derivative is small). 

On the other hand, the inhomogeneous cooling effect of the IR opacity is much larger. If we fix the visible opacity and only vary the IR opacity, the dependence of $F_{\rm int}$ (\Cref{fig:Firesponse}D) is similar to that when changing both visible and IR opacity (\Cref{fig:Firesponse}B). Thus, the inhomogeneity of incident stellar flux and/or atmospheric opacity would generally enhance the observed internal heat flux at the top of the atmosphere and interior cooling on giant planets.

\subsection{Internal Heat Flux Change with Increasing Inhomogeneity}

Similar to the case of terrestrial planets, we demonstrated the simplest inhomogeneous examples for giant planets using only two atmospheric columns with different incident stellar flux and atmospheric opacity. We then varied the contrast between the two columns and compared the mean internal heat flux with the homogeneous case with the mean incident stellar flux and opacity.

We assumed $D$=2, $\beta\sim 0.76$, $K=0.2$, and two different values of $\alpha=0.5$ and 5. The RCE temperature profiles are shown in \Cref{fscaling}A. As in previous studies (i.e., see the RCE solution in \citealt{robinsonAnalyticRadiativeConvectiveModel2012} or the RE solutions in \citealt{guillotRadiativeEquilibriumIrradiated2010}), a temperature inversion is developed in the radiative zone if $\alpha$ is large, resulting in a hotter atmosphere with a larger stellar flux. The temperature in the homogeneous column is located between the two inhomogeneous columns. For the case without thermal inversion ($\alpha=0.5$), the stellar flux can penetrate deeper into the atmosphere, and the RCB is lower. Consequently, the outgoing internal heat flux is smaller than the case with a larger $\alpha$ as shown in \Cref{fig:Firesponse}A.

We plotted the internal flux ratio as a function of the incident stellar flux contrast between the two columns in \Cref{fscaling}C. For all cases, with and without thermal inversion, the flux ratio is larger than unity, suggesting that the internal heat flux is always larger in the inhomogeneous columns. The ratio monotonically increases with the column contrast (\Cref{fscaling}C) and reaches above a factor of 2 in the cases we simulated here. In other words, with significant inhomogeneity in incident stellar flux, the atmosphere could emit more than twice the internal heat flux in a uniform atmosphere with an average incident stellar flux.

For the effect of inhomogeneous opacity, \Cref{fscaling}B shows the temperature profiles in the normalized pressure coordinate as in the terrestrial cases. With a larger $\kappa$, the atmosphere has a larger opacity at a given pressure level. Even though the RCB pressure levels in these cases do not change much (\Cref{fscaling}B), their corresponding optical depths are very different, and so are the internal heat fluxes. The internal flux ratio between the inhomogeneous columns and the homogeneous column is always larger than unity and increases drastically with the column opacity contrast (\Cref{fscaling}D). As the contrast increases, the ratio shows a steeper increase in the irradiated atmosphere than the self-luminous one, reaching over a factor of 50 larger than the uniform column if the opacity contrast exceeds 100. A factor of 100 difference in opacity might be exaggerated in the deep atmosphere where the species are expected to be well mixed and the $\mathrm{H_2}$-$\mathrm{H_2}$ CIA contributes the most to the opacity. But a large opacity contrast is possible in the upper atmosphere, especially if some short-lived photochemical species dominate the opacity, the horizontal difference of which could differ by orders of magnitude (e.g., \citealt{mosesChemicalVariationAltitude2021}). The patchy clouds could also produce a large opacity contrast. Unfortunately, limited by our analytical framework, we assume the opacity contrast is the same throughout the atmospheric column from top to bottom. Numerical simulations should investigate a more realistic situation. But it seems significant that opacity contrast can impact the mean internal heat flux on both irradiated and self-luminous bodies and tends to increase the outgoing planetary heat flow. 

We varied the incident stellar flux and opacity contrast between the two columns by a factor of 0.05 to 20 to explore the parameter space. We adopted $D=2$ and $\alpha=1$ with the reference incident stellar flux of 1 and the reference opacity scale of 1 (corresponding to $K=0.2$). The ratio of the averaged flux from the two inhomogeneous columns to the homogeneous case is shown in \Cref{fcontour}. The internal flux ratio is larger than unity, increasing towards high contrast cases. Similar to the terrestrial cases (\Cref{tecontour}), the pattern seems centrosymmetric about the reference point. The combination of the incident stellar flux and opacity contrast can enhance or compensate for the effects of each other. But the behavior is different from the terrestrial case. For giant planets, the internal heat flux ratio reaches the maximum if one column has a higher incident stellar flux and opacity than the other, reaching above 10 for the extremely inhomogeneous case when the contrast is 20. If one column has a higher stellar flux but a lower opacity, the internal flux enhancement is smaller. The reasons behind the differences between terrestrial and giant planet cases are yet to be explored. However, just as with terrestrial instances, the two-column examples in the giant planet case can be extrapolated to understand the average scenario involving multiple columns. The outgoing internal heat flux is expected to increase in the multi-column case, although the extent of this increase would depend on the degree of inhomogeneity.

\section{Conclusion and Discussion} \label{consec}

As a fundamental property of a planet, inhomogeneity has often been overlooked in previous heat flow analyses. Our proposed principle states that inhomogeneities in the surface and atmosphere tend to accelerate planetary cooling. This is based on Jensen's inequality and the nonlinear responses of surface temperature and outgoing heat flux to changes in incoming stellar flux and atmospheric opacity. We conducted proof-of-concept tests using semi-gray analytical models and have found that this principle holds for a wide range of parameters. Key findings include:

1. The Planck blackbody law is a convex function of temperature. The temporally and spatially mean surface temperature of an airless planet is lower than its equilibrium temperature under the global energy balance.

2. For terrestrial planets with atmospheres, we examine the dependence of the surface temperature on individual model parameters in the system while holding other parameters constant. The surface temperature increases with increasing incident stellar flux and IR opacity but decreases with increasing visible opacity. The individual response of the surface temperature to changes in incident stellar flux, visible, and IR opacities is concave. This means that a planet with inhomogeneously distributed incident stellar flux and atmospheric opacity has a lower mean surface temperature than a uniform planet with the same total incident stellar flux and opacity.

3. For giant planets, the dependence of the outgoing internal heat flux on individual model parameters in the system was also analyzed. The outgoing interior heat flux decreases with increased incident stellar flux and IR opacity but increases with increased visible opacity. The individual response of the outgoing radiative flux to changes in incident stellar flux and IR opacity is convex, while the response to changes in visible opacity is concave. It implies that, given the same internal entropy, inhomogeneity in incident stellar flux and IR opacity enhances the cooling of the planet's interior compared to a uniform case. Inhomogeneity in visible opacity reduces the interior cooling, but the effect is much smaller than the enhancement caused by IR opacity inhomogeneity. Overall, inhomogeneity in the atmosphere accelerates the cooling of the interior of giant planets.

4. The effect of inhomogeneity on surface temperature and outgoing flux can be quantified by comparing two inhomogeneous columns to a uniform column. The decrease in surface temperature or increase in outgoing flux is more pronounced as the contrast between the two columns increases. When incident stellar flux and opacity contrast are both varied, the surface temperature on terrestrial planets can be reduced by more than 20\%, and the outgoing radiative flux on giant planets can be increased by more than an order of magnitude compared to a uniform case. 

5. The temperature reduction and internal heat flux increase in the two-column tests can be extrapolated to shed light on scenarios involving multiple columns, where the inhomogeneity effect is also expected to persist, as suggested by Jensen's inequality.

Given the current focus on global-mean surface temperature as a critical factor in planetary habitability (including in the climate studies on Earth) and global-mean evolution models of giant planets, it is important to highlight the effect of inhomogeneity. 3D planets naturally harbor the inhomogeneity of incident stellar flux. Planets with large albedo contrast, high obliquity, or eccentric orbits also possess significant surface and atmospheric inhomogeneity. In those situations, not just the equilibrium temperature but also the mean surface temperature calculated using 1D RCE models cannot accurately reflect the planet's habitability. This argument expands upon the critique presented in \cite{leconte3DClimateModeling2013} to include a wider range of planets beyond just synchronized terrestrial planets.

Incorporating the effect of inhomogeneity could improve our understanding of the evolution of giant planets. The outgoing cooling flux could be significantly different in the presence of large inhomogeneity in the weather layer. Our findings are consistent with previous research on tidally locked exoplanets (\citealt{guillotEvolution51Pegasus2002,budajDayNightSide2012,spiegelThermalProcessesGoverning2013,rauscherINFLUENCEDIFFERENTIALIRRADIATION2014}), but have broader implications for all giant planets with spatially inhomogeneous and temporally varying atmospheres, such as those with high obliquities and in eccentric orbits. This also suggests that rotational states could impact temperature distribution and the cooling rate of the interior. In the next paper (\citealt{zhangInhomogeneityEffectII2023}), we will examine the effects of rotational and orbital states on interior cooling. In our third paper in this series (\citealt{zhangInhomogeneityEffectIII2023}), we will conduct numerical simulations on tidally locked exoplanets to shed light on the heating and cooling mechanisms of hot Jupiters.

Inhomogeneity in IR opacity can affect the outgoing heat flow of both irradiated and self-luminous planets. This finding has implications for planetary evolution, as well as the emission flux calculation from 2D/3D atmospheric models with coarse grids, because sub-grid-scale inhomogeneity may enhance the outgoing infrared flux. We should pay close attention to the resolution-dependence of total outgoing flux in simulations of cloudy atmospheres, as patchy clouds often have very high opacity contrast (e.g., \citealt{zhangAtmosphericRegimesTrends2020, tanAtmosphericCirculationBrown2021a, tanAtmosphericCirculationBrown2021}). 

It is important to examine the simplifications in our analytical RCE framework closely. The effect of inhomogeneities in the incident stellar flux appears robust, as the flux distribution should follow the geometry of a 3D sphere and the rotational and orbital configurations. However, the effect of opacity inhomogeneity is more complex. First, we have only considered the simplest case of inhomogeneity, using a semi-gray, cloud-free atmosphere in this study. We have assumed that infrared opacity is a power law function of the pressure and that the visible opacity is proportional to IR opacity. But realistic atmospheres are non-gray with cloud and haze scattering. Second, limited by our analytical framework, inhomogeneity was only introduced by scaling the entire opacity profile from top to bottom. In real atmospheres, the local opacity often varies at a certain level due to local chemistry, dynamics, and cloud formation, leading to the shape change in the opacity profiles. Moreover, opacity does not solely depend on the abundance of the absorbers but also on the absorption coefficient of the species, which could be a strong function of local temperature. Even though the total mass of the absorber might be conserved during the homogenization processes as long as the chemistry is not important, the total opacity might not be if a strong temperature variation exists. 

Although there is no doubt that the opacity inhomogeneity would significantly alter planetary cooling, further research with detailed opacity treatment is needed to revisit our proof-of-concept calculations on the cooling (or even heating) effects of opacity inhomogeneity in more complex situations. This requires numerical radiative transfer simulations and is beyond the analytical framework of this study.

Additional atmospheric processes should also be considered beyond our simple 1D framework. The real atmosphere is not in RCE, where 3D atmospheric dynamics can homogenize temperature and opacity. On temperate terrestrial planets where the weak temperature gradient approximation holds, the free atmosphere is roughly homogeneous (e.g., \citealt{sobelWeakTemperatureGradient2001}). In this case, the effect of inhomogeneity might not be substantial. However, for tidally locked exoplanets with extreme temperature differences between the day and night sides, the near-surface boundary layer may not be homogeneous (e.g., \citealt{wordsworthAtmosphericHeatRedistribution2015, kollTemperatureStructureAtmospheric2016}), and the difference between 1D and 3D dry models is evident  (\citealt{leconte3DClimateModeling2013}). Moisture in the atmosphere may also cause a more linear response to surface emission, as seen in long-term Earth climate data (\citealt{kollEarthOutgoingLongwave2018}), which would reduce the importance of spatial and temporal inhomogeneity in the atmosphere. 

It should also be noted that the inhomogeneity effect, while significant, does not account for all weather effects on planetary heat transport that are neglected in the current 1D framework, as other processes, such as heat advection by large-scale circulation, eddy heat mixing, and latent heat transport may also be important. In our third paper (\citealt{zhangInhomogeneityEffectIII2023}), we will use 3D realistic radiative-dynamical simulations to evaluate the impact of weather on the interior cooling of tidally locked gas giants.

Finally, investigating the impact of inhomogeneity emphasizes the importance of statistical distributions of planetary variables on mean properties. This raises the question of how to measure and compare statistical distributions among different planets. In this study, we have qualitatively compared inhomogeneous cases to uniform cases to demonstrate the bias in current 1D or low-resolution models using Jensen's inequality. However, quantifying the difference among inhomogeneous cases requires a better understanding of estimating the size of the Jensen gap (e.g., \citealt{liaoSharpeningJensenInequality2019}), which is a currently unsolved problem in statistics. Because realistic planets all have some degree of inhomogeneity, we will need to consider how to measure the level of inhomogeneity and compare two inhomogeneous planets. In Appendix \ref{app:convex}, we have tentatively introduced a way to compare two inhomogeneous cases with two data points. A general theory will be further developed in our next paper (\citealt{zhangInhomogeneityEffectII2023}).
 
\section{acknowledgments}

We acknowledge the reviewer for their careful reading and helpful comments. X.Z. is supported by NASA Exoplanet Research Grant 80NSSC22K0236 and the NASA Interdisciplinary Consortia for Astrobiology Research (ICAR) grant 80NSSC21K0597. This work was inspired by the last in-person conversation with Dr. Adam P. Showman (1969-2020) about the heat flow on tidally locked exoplanets at the AGU Fall Meeting 2019. We also acknowledge the assistance provided by the OpenAI language model.

%% To help institutions obtain information on the effectiveness of their 
%% telescopes the AAS Journals has created a group of keywords for telescope 
%% facilities.
%
%% Following the acknowledgments section, use the following syntax and the
%% \facility{} or \facilities{} macros to list the keywords of facilities used 
%% in the research for the paper.  Each keyword is check against the master 
%% list during copy editing.  Individual instruments can be provided in 
%% parentheses, after the keyword, but they are not verified.

\vspace{5mm}
%\facilities{HST(STIS), Swift(XRT and UVOT), AAVSO, CTIO:1.3m,
%CTIO:1.5m,CXO}

%% Similar to \facility{}, there is the optional \software command to allow 
%% authors a place to specify which programs were used during the creation of 
%% the manuscript. Authors should list each code and include either a
%% citation or url to the code inside ()s when available.

\iffalse

\software{Numpy \citep{NumpyCitation},
          SciPy \citep{SciPyCitation}, 
          Matplotlib \citep{MatplotlibCitation},
          Cartopy \citep{CartopyCitation},
          %astropy \citep{2013A&A...558A..33A,2018AJ....156..123A},  
          windspharm \citep{dawson2016windspharm},
          Bibmanager \url{https://bibmanager.readthedocs.io/en/latest/index.html}
          }

\fi
\appendix

\section{Proof of the Convexity of the Planck Function}
\label{app:planck}
\textit{Proposition I.} The Planck blackbody law as a function of frequency is written as:
\begin{equation}
B_\nu(T) = \frac{2h\nu^3}{c^2} \frac{1}{e^{\frac{h\nu}{k_BT}} - 1},
\end{equation}
where $B_\nu(T)$ is the radiance at frequency $\nu$ and temperature $T$, $h$ is the Planck constant, $c$ is the speed of light in a vacuum, $k_B$ is the Boltzmann constant, and $T$ is the temperature of the blackbody. $B_\nu(T)$ is a convex function of temperature $T$:
\begin{equation}
\frac{\partial^2 B}{\partial T^2} > 0.
\end{equation}

\textit{Proof.} The second derivative of $B_\nu(T)$ with respect to $T$ is:
\begin{equation}
\frac{\partial^2 B}{\partial T^2} = \frac{2k_B^2\nu}{hc^2} \frac{x^3e^x\left(xe^x-2e^x+x+2\right)}{(e^x-1)^3},
\end{equation}
where $x=h\nu/k_BT$. To prove that $\partial^2 B/\partial T^2 > 0$, we must show that the function $G(x)$ defined as:
\begin{equation}
G(x) = xe^x-2e^x+x+2,
\end{equation}
is always positive for any positive $x$. We can accomplish this by demonstrating that $G(x)$ is a convex function for positive $x$. First, we find the derivative of $G(x)$:
\begin{equation}
G^\prime(x) = xe^x-e^x+1,
\end{equation}
and the second derivative of $G(x)$:
\begin{equation}
G^{\prime\prime}(x) = xe^x+1.
\end{equation}

It is clear that $G^{\prime\prime}(x)$ is strictly positive for all positive $x$. Because $G^\prime(0) = 0$ and $G^{\prime\prime}(x) > 0$ for all positive $x$, it follows that $G^\prime(x) > 0$ for all positive $x$. Similarly, because $G(0) = 0$ and $G^\prime(x) > 0$ for all positive $x$, it follows that $G(x) > 0$ for all positive $x$. Thus, $\partial^2 B/\partial T^2 > 0$, which proves that $B_\nu(T)$ is a convex function of temperature $T$.

\section{Alternative Derivation of the Two-Stream Radiative Diffusion Equation}
\label{app1}

We start from the radiative transfer equation in the pure-absorption limit:
\begin{equation}
\mu \frac{dI(\tau, \mu, \omega)}{d\tau} = I(\tau, \mu, \omega) - J(\tau, \mu, \omega),
\label{rt1}
\end{equation}
where $\tau$ is the optical depth, $I$ is the intensity, $J$ is the source function, $\mu$ is the cosine of the zenith angle, and $\omega$ is the azimuthal angle.

\Cref{rt1} can be transformed into an integral form (see, e.g., \citealt{goodyAtmosphericRadiationTheoretical1995} p. 47):
\begin{equation}
\begin{split}
I(\tau, \mu, \omega) &= I_s(\tau_{\rm s}, \mu, \omega)e^{-(\tau_{\rm s} - \tau)/\mu} + \int_\tau^{\tau_{\rm s}}J(\tau, \mu, \omega)e^{-(t - \tau)/\mu} \frac{dt}{\mu} ~~~~~~~ \mathrm{for}~ 0 < \mu \leq 1 \\
I(\tau, \mu, \omega) &= -\int_0^{\tau} J(\tau, \mu, \omega)e^{-(t - \tau)/\mu} \frac{dt}{\mu} ~~~~~~~~~~~~~~~~~~~~~~~~~~~~~~~~~ \mathrm{for}~ -1 \leq \mu < 0.
\end{split}
\label{rt}
\end{equation}
Here, $I_s$ is the lower boundary upward stream at optical depth $\tau_{\rm s}$. We have assumed no downward thermal radiation from the top of the atmosphere. Under the assumption of local thermodynamical equilibrium (LTE), the source function $J$ can be replaced by the blackbody source function $B(\tau)$.

By taking the integration of \Cref{rt} over the zenith and azimuth angles, we obtain the upward flux $F^+$ and downward flux $F^-$, respectively:
\begin{equation}
\begin{split}
F^+(\tau) &= 2\pi B(\tau_{\rm s})E_3(\tau_{\rm s} - \tau) + 2\pi \int_\tau^{\tau_{\rm s}} B(t)E_2(t - \tau) dt \\
F^-(\tau) &= -2\pi \int_0^{\tau} B(t)E_2(\tau - t) dt,
\end{split}
\label{f0}
\end{equation}
where $E_n(x) = \int_1^{\infty} \frac{e^{-wx}}{w^n} dw$ is the exponential integral function.

By taking the derivative of the flux for the optical depth $\tau$ using the Leibniz rule, we obtain:
\begin{equation}
\begin{split}
\frac{dF^+}{d\tau}&=2\pi B(\tau_{\rm s})E_2(\tau_{\rm s}-\tau) - 2\pi B(\tau) E_2(0) + 2\pi\int_\tau^{\tau_{\rm s}} B(t)E_1(t-\tau)dt \\
\frac{dF^-}{d\tau}&= -2\pi B(\tau) E_2(0) + 2\pi\int_0^\tau B(t)E_1(\tau-t)dt,
\end{split}
\label{f1}
\end{equation}
where $E_2(0)=1$. To obtain a diffusional form, the traditional treatment directly approximates the transmission function $E_3(x)\sim e^{-Dx}/2$ (e.g., \citealt{armstrongTheoryDiffusivityFactor1968, goodyAtmosphericRadiationTheoretical1995}). $D$ is called the diffusivity factor. Accordingly, $E_2(x)\sim De^{-Dx}/2$ and $E_1(x)\sim D^2e^{-Dx}/2$. Under this assumption, the radiative diffusion equation becomes:
\begin{equation}
\frac{d^2F}{d\tau^2}=D^2F - 2\pi D \frac{dB}{d\tau}.
\label{olddrt}
\end{equation}

Here we introduce another treatment. The essence of the traditional method is to link the flux derivative back to the flux. Instead of approximating $E_n(x)$, we approximate their ratios such that:
\begin{equation}
\frac{E_1(x)}{E_2(x)}\approx\frac{E_2(x)}{E_3(x)}\approx D.
\label{appx}
\end{equation}
With this assumption, the upward and downward fluxes can be written as:
\begin{equation}
\begin{split}
\frac{dF^+}{d\tau}&= -2\pi B(\tau) + DF^+ \\
\frac{dF^-}{d\tau}&= -2\pi B(\tau) - DF^-.
\end{split}
\label{fluxeq}
\end{equation}

From the two above equations, we can obtain a new radiative diffusion equation for the net flux $F=F^++F^-$:
\begin{equation}
\frac{d^2F}{d\tau^2}=D^2F - 4\pi \frac{dB}{d\tau}.
\label{drt}
\end{equation}

\begin{figure}
\centering
\includegraphics[width=0.45\textwidth]{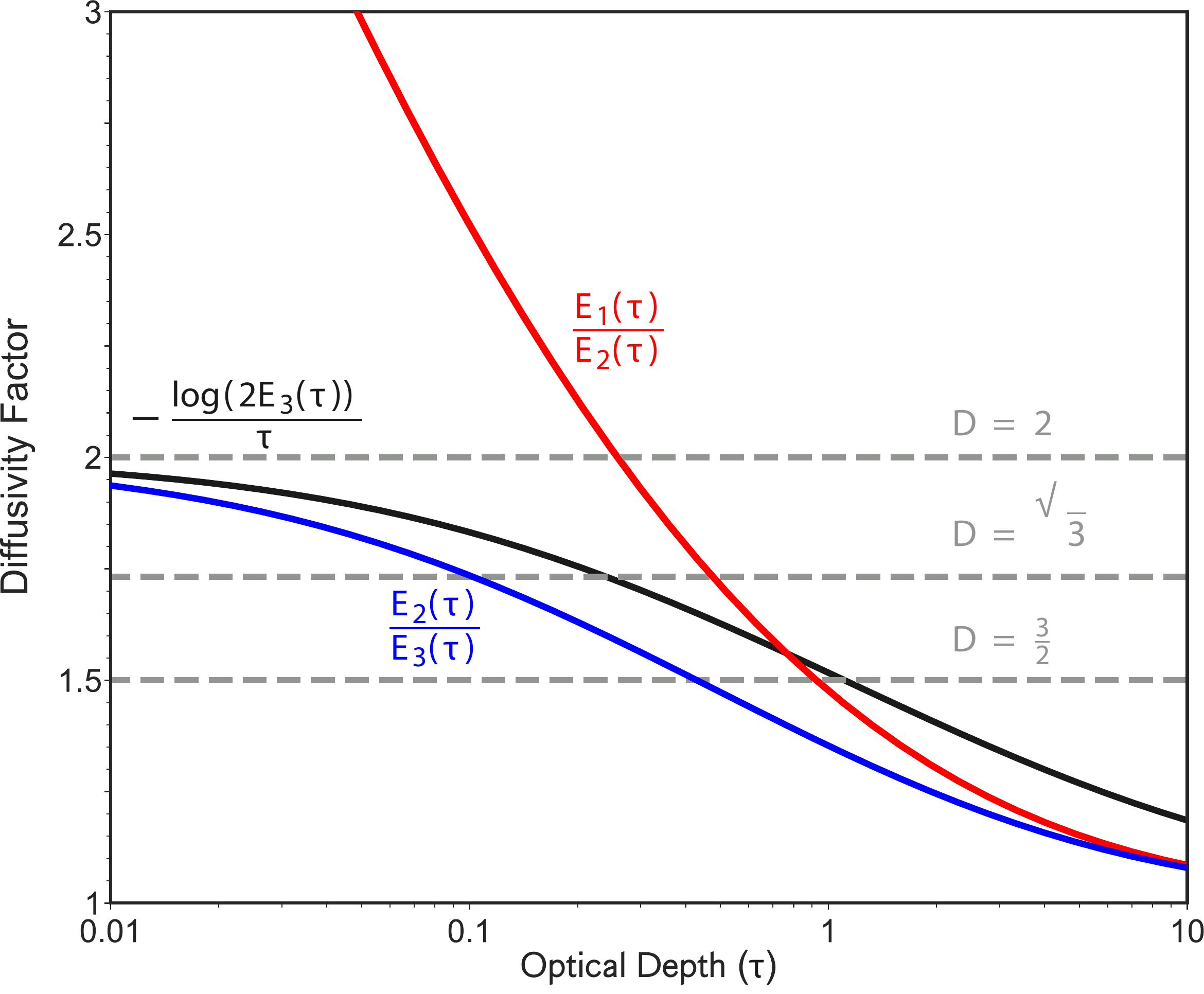}
\caption{Diffusivity factor as a function of optical depth from different theories: $D=-\log(2E_3(x))/x$ (black) in \cite{armstrongTheoryDiffusivityFactor1968}, $D=E_1(x)/E_2(x)$ (red), and $D=E_2(x)/E_3(x)$ (blue) in this work. Typical constant values of $D$ are used in practice (gray dashed). }
\label{dfac}
\end{figure}

This new radiative diffusive equation slightly differs from the traditional form \Cref{olddrt} where the prefactor in front of the second term on the right-hand side is $2\pi D$ instead of $4\pi$. This is because the traditional treatment assumes $E_2(x)\sim De^{-Dx}/2$ such that $E_2(0)\sim D/2$, but we keep $E_2(0)=1$ in our form.

The choice of the value of the diffusivity factor $D$ is an art. To diagnose how the good is the assumption of a constant $D$, we plotted $D=-\log(2E_3(x))/x$ for the traditional approximation and $D=E_1(x)/E_2(x)$ and $E_2(x)/E_3(x)$ in the new approximation (Figure \ref{dfac}). When $x$ increases from 0.01 to 1, both $-\log(2E_3(x))/x$ and $E_1(x)/E_2(x)$ decrease by a factor of two but $E_2(x)/E_3(x)$ decreases by more than a factor of three. We have also assumed $E_1(x)/E_2(x)=E_2(x)/E_3(x)$ (Equation \ref{appx}) but the two curves only converge when the optical depth is high. The differences between the traditional diffusivity factor ($\log(2E_3(x))/x$) and our diffusivity factors ($E_1(x)/E_2(x)$ and $E_2(x)/E_3(x)$) are usually smaller than 10\% when the optical depth is larger than about 0.35. Because $D$ is a constant, it is often empirically determined in the traditional diffusivity equation, such as $D=1.66$ (\citealt{rodgersComputationInfraredCooling1966, armstrongTheoryDiffusivityFactor1968}), $D = 3/2$ (\citealt{weaverDeductionsSimpleClimate1995}), and $D = 2$ for the hemi-isotropic approximation (\citealt{pierrehumbertPrinciplesPlanetaryClimate2010}). Some more complicated forms of $D$ as a function of the optical depth have been introduced (e.g., \citealt{zhaoAccurateApproximationDiffusivity2013}).

Unlike the equation under the Eddington approximation, the traditional diffusivity equation cannot simultaneously satisfy the optionally-thin and thick limits (\citealt{goodyAtmosphericRadiationTheoretical1995}, p. 63). Interestingly, our radiative diffusion \Cref{drt} is consistent with the equation using the Eddington approximation (e.g., \citealt{hubenyPossibleBifurcationAtmospheres2003, guillotRadiativeEquilibriumIrradiated2010}) if we use $D=\sqrt{3}$. It also converges to the equation under the hemi-isotropic diffusive equation if $D=2$. \Cref{drt} gives us some flexibility to explore the system's behavior under previous assumptions in one framework.

In the gray limit, the blackbody source function $B=\sigma T^4/\pi$ where $\sigma$ is the Stefan-Boltzmann constant, and $T$ is the local temperature. If we normalize the temperature by a reference temperature $T_0$ and normalize the flux by $4\sigma T_0^4$, \Cref{drt} becomes:
\begin{equation}
\frac{d^2F}{d\tau^2}=D^2F - \frac{dT^4}{d\tau}.
\label{drt0}
\end{equation}

\section{Upward and Downward Fluxes for Generic Radiative-Equilibrium Atmospheres}
\label{app2}

The upward and downward fluxes can be solved in \Cref{fluxeq}. Multiplying $e^{-Dt}$ to both sides and integrating the equations, we obtain:
\begin{equation}
\begin{split}
F^+(\tau)&=\pi B(\tau_{\rm s})e^{D(\tau-\tau_{\rm s})}+2\pi\int_\tau^{\tau_{\rm s}} B(t)e^{-D(t-\tau)}dt \\
F^-(\tau)&=-2\pi\int_0^\tau B(t)e^{-D(\tau-t)}dt.
\end{split}
\label{fluxsol}
\end{equation}
The fluxes are slightly different from the traditional form (e.g., see Equation (A6) in \citealt{robinsonAnalyticRadiativeConvectiveModel2012}). Using the normalized radiative-equilibrium temperature structure \Cref{trad} with the expression of $S(\tau)$ in \Cref{srceq}, we can obtain the downward flux $F^-(\tau)$ normalized  by $4\sigma T_{0}^4$:
 \begin{equation}
 \begin{split}
 F_{\rm rad}^-(\tau)&=-\frac{1}{2}\int_0^\tau \big[S(t)+(D+D^2t)F_{\rm int}\big]e^{-D(\tau-t)}dt\\
 &=-\frac{1}{2}\int_0^\tau \bigg[\frac{\partial F_{\rm v}}{\partial t}-D^2\int_0^t F_{\rm v}dt-DF_{\rm v}(0)+(D+D^2t)F_{\rm int}\bigg]e^{-D(\tau-t)}dt.
\end{split}
\end{equation}
Note that:
\begin{equation}
\int(\frac{\partial F_{\rm v}}{\partial t}+DF_{\rm v})e^{Dt}dt=\int\frac{\partial (F_{\rm v}e^{Dt})}{\partial t}dt=F_{\rm v}e^{Dt}.
\end{equation}
We can rewrite the integral of $F^-(\tau)$ as:
 \begin{equation}
 \begin{split}
 F_{\rm rad}^-(\tau)&=-\frac{1}{2}\int_0^\tau \bigg[(\frac{\partial F_{\rm v}}{\partial t}+DF_{\rm v})-D(F_{\rm v}+D\int F_{\rm v}dt)+D\big[(D\int F_{\rm v}dt-F_{\rm v})\Big|_{\tau=0}+F_{\rm int}\big]+D^2F_{\rm int}t\bigg]e^{-D(\tau-t)}dt\\
&=-\frac{1}{2}\left[F_{\rm v}(\tau)-D\int_0^\tau F_{\rm v}(\tau)d\tau-F_{\rm v}(0)+D\tau F_{\rm int}\right].
 \end{split}
\end{equation}
Rearranging it gives:
\begin{equation}
 F_{\rm rad}^-(\tau)=-\frac{1}{2}\left[\frac{S(\tau)}{D}-\frac{1}{D}\frac{\partial F_{\rm v}}{\partial\tau}+F_{\rm v}(\tau)+D\tau F_{\rm int}\right].\label{dnfeqn}
\end{equation}
Similarly, we can calculate the normalized $F_{\rm rad}^+(\tau)$:
\begin{equation}
 \begin{split}
    F_{\rm rad}^+(\tau)&=\frac{T_{\rm s}^4e^{D(\tau-\tau_{\rm s})}}{4}+\frac{1}{2}\int_\tau^{\tau_{\rm s}} \big[S(t)+(D+D^2t)F_{\rm int}\big]e^{-D(t-\tau)}dt \\
    &=\frac{T_{\rm s}^4e^{D(\tau-\tau_{\rm s})}}{4}-\frac{1}{2}e^{D(\tau-\tau_{\rm s})}\left[\frac{S(\tau_{\rm s})}{D}-\frac{1}{D}\frac{\partial F_{\rm v}}{\partial\tau}\Big|_{\tau_{\rm s}}-F_{\rm v}(\tau_{\rm s})+(2+D\tau_{\rm s})F_{\rm int}\right] \\
   &+\frac{1}{2}\left[\frac{S(\tau)}{D}-\frac{1}{D}\frac{\partial F_{\rm v}}{\partial\tau}-F_{\rm v}(\tau)+(2+D\tau)F_{\rm int}\right]\\
   &=\left[\frac{T_{\rm s}^4}{4}+F^-(\tau_{\rm s})+F_{\rm v}(\tau_{\rm s})-F_{\rm int}\right]e^{D(\tau-\tau_{\rm s})}+\frac{1}{2}\left[\frac{S(\tau)}{D}-\frac{1}{D}\frac{\partial F_{\rm v}}{\partial\tau}-F_{\rm v}(\tau)+(2+D\tau)F_{\rm int}\right].
\end{split}
\end{equation}
Note that the surface albedo has been implicitly included in $F_{\rm v}$. The first term on the right-hand side should vanish due to the energy conservation at the surface under radiative equilibrium. Also, $e^{D(\tau-\tau_{\rm s})}$ vanishes as $\tau_{\rm s} \rightarrow \infty$ on giant planets. $F_{\rm rad}^+(\tau)$ thus becomes:
\begin{equation}
    F_{\rm rad}^+(\tau)=\frac{1}{2}\left[\frac{S(\tau)}{D}-\frac{1}{D}\frac{\partial F_{\rm v}}{\partial\tau}-F_{\rm v}(\tau)+(2+D\tau)F_{\rm int}\right].\label{upfeqn}
\end{equation}
One can show that the solutions of $F_{\rm rad}^+(\tau)$ and $F_{\rm rad}^-(\tau)$ satisfy the radiative equilibrium at every atmospheric level, i.e.,  $F_{\rm rad}^+(\tau)+F_{\rm rad}^-(\tau)=-F_{\rm v}(\tau)+F_{\rm int}$. 

For terrestrial planets, $F_{\rm int}\approx 0$. The radiative equilibrium condition at the surface gives the surface temperature:
\begin{equation}
    \frac{T_{\rm s}^4}{4}=-F_{\rm rad}^-(\tau_{\rm s})-F_{\rm v}(\tau_{\rm s})=\frac{1}{2}\left[\frac{S(\tau_{\rm s})}{D}-\frac{1}{D}\frac{\partial F_{\rm v}}{\partial\tau}\Big|_{\tau_{\rm s}}-F_{\rm v}(\tau_{\rm s})\right].
\end{equation}
This equation implies a discontinuity between the surface temperature $T_{\rm s}^4$ and the radiative-equilibrium temperature $S(\tau_{\rm s})$ of the overlying atmosphere in direct contact with the surface. To make the analysis simpler, we consider a simple attenuated stellar flux following Beer's law, where $F_{\rm v}(\tau)=-F_{\odot}e^{-\alpha\tau}$ and $\alpha=\kappa_v/\kappa$ is the ratio of the visible opacity to the IR opacity under the semi-gray approximation. The temperature discontinuity can then be expressed as:
\begin{equation}
    T_{\rm s}^4-S(\tau_{\rm s})=\frac{1}{D}\left[(2-D)S(\tau_{\rm s})+2(D-\alpha)F_{\odot}e^{-\alpha\tau_{\rm s}}\right].\label{surft}
\end{equation}
If the hemi-isotropic approximation is used with a value of $D=2$, and $\alpha < D$, the surface temperature will be warmer than the temperature of the overlying air. In this case, near-surface convection will be triggered. On the other hand, if $\alpha > D$, the surface temperature may be cooler, and the system will be stable. As shown in Appendix \ref{app3}, a value of $\alpha > D$ also implies a stably stratified atmosphere at all levels according to the Schwarzschild criterion. In this situation, in addition to radiation, thermal conduction at the surface may occur to smooth out the temperature discontinuity. Thin atmospheres on icy bodies like Pluto and Triton provide an example of this scenario (e.g., see \citealt{zhangHazeHeatsPluto2017,strobelComparativePlanetaryNitrogen2017,gladstoneNewHorizonsObservations2019}).

\section{Schwarzschild criterion for convective stability in a radiative-equilibrium atmosphere}
\label{app3}
The Schwarzschild criterion (Equation \ref{sch}) sets a sufficient condition for the convective instability. In other words, it is the necessary condition of convective stability in the radiative zone:
\begin{equation}
\frac{d\ln T_{\rm rad}^4}{d\ln \tau}<\beta.
\end{equation}
Note that the above criterion has assumed a power-law dependence of atmospheric opacity on the pressure but not on temperature (see Equation \ref{adt}). Also, $\beta$ could vary in real atmospheres but is assumed constant in this study. By inserting the radiative-equilibrium temperature structure (given by Equation \ref{trad}) and the expression for the source function $S(\tau)$ (given by Equation \ref{srceq}), we can rewrite the Schwarzschild criterion as:
\begin{equation}
\frac{\frac{\partial^2 F_{\rm v}}{\partial\tau^2}\tau-D^2F_{\rm v}(\tau)\tau+D^2F_{\rm int}\tau}{\frac{\partial F_{\rm v}}{\partial\tau}-D^2\int_0^\tau F_{\rm v}d\tau-DF_{\rm v}(0)+(D+D^2\tau)F_{\rm int}}<\beta.\label{scrterion}
\end{equation}

In the deep atmosphere of a giant planet, the incident stellar flux $F_{\rm v}$ is negligible but $F_{\rm int}$ is not zero. \Cref{scrterion} becomes:
\begin{equation}
\frac{1}{1+(D\tau)^{-1}}<\beta.
\end{equation}
In the limit of infinitely large $\tau$, the stability criterion suggests that $\beta > 1$ when $\tau\rightarrow\infty$. This is because the temperature structure $T_{\rm rad}(\tau)$ (Equation \ref{trad}) is expected to scale as $T_{\rm rad}^4(\tau)\propto\tau$ as $\tau\rightarrow\infty$ and thus the gradient of the radiative energy ($d\ln T_{\rm rad}^4/d\ln \tau$) is unity. Based on the Schwarzschild criterion, $\beta > 1$ is required to ensure the temperature gradient under radiative equilibrium would not exceed the adiabatic lapse rate. Because $\beta=4\zeta(\gamma-1)/n\gamma$, if we assume a diatomic gas with $\gamma=7/5$ and $\zeta=1$, $\beta > 1$ requires $n<8/7$. Therefore, if $n<8/7$, the atmospheres may not develop a convective zone. However, as noted in the main text, values of $n$ are generally larger than unity in realistic atmospheres, so convective instability will likely occur even if the interior heat flux is small but non-zero.

For terrestrial planets, the interior heat flux $F_{\rm int}=0$ and the stellar flux determines the radiative-equilibrium temperature structure in the atmosphere. To simplify the discussion, we assume a simple direct beam $F_{\rm v}(\tau)=-F_{\odot}e^{-\alpha\tau}$, in which case \Cref{scrterion} becomes:
\begin{equation}
\frac{(D^2-\alpha^2)e^{-\alpha\tau}\alpha\tau}{(\alpha^2-D^2)e^{-\alpha\tau}+D^2+D\alpha}<\beta.\label{sc2}
\end{equation}

This equation is equivalent to Equation (33) in \cite{robinsonAnalyticRadiativeConvectiveModel2012} (note that they used $k$ instead of $\alpha$). They found that the atmosphere is against convection if $\alpha>D$ (see also a similar argument in \citealt{pierrehumbertPrinciplesPlanetaryClimate2010}, p.212). \cite{robinsonAnalyticRadiativeConvectiveModel2012} also noted that the lapse rate (the left-hand side of the above equation) has a maximum value and computed several cases in their Figure 4. Here, we show that the maximum value can be analytically derived to provide a criterion for convective stability in the radiative zone.

After some algebra, \Cref{sc2} can be rewritten as:
\begin{equation}
\frac{\alpha}{D}>1-e^{\alpha\tau}\left(1+\frac{\alpha\tau}{\beta}\right)^{-1}.\label{ch3}
\end{equation}
The maximum value of the right-hand side is $1-\beta e^{1-\beta}$ at $\tau=(1-\beta)/\alpha$. Thus, the atmosphere might be stable against convection as long as: 
\begin{equation}
\frac{\alpha}{D}>1-\beta e^{1-\beta}.
\end{equation}
This is a stronger condition than the previous criterion $\alpha>D$ in \citet{robinsonAnalyticRadiativeConvectiveModel2012}. Given the values of $\alpha$, $\beta$, $D$, and $\tau_{\rm s}$, the atmosphere must develop a convective zone if both of the following conditions are met:
\begin{equation}
\begin{split}
&\frac{\alpha}{D}<1-\beta e^{1-\beta}, \\
&\tau_{sc}<\tau_{s},
\end{split}
\end{equation} 
where $\tau_{sc}$ is a critical value given by:
\begin{equation}
e^{\alpha\tau_{sc}}\left(1+\frac{\alpha\tau_{sc}}{\beta}\right)^{-1}=1-\frac{\alpha}{D}.
\end{equation}
$\tau_{sc}$ marks the location of RCB under the Schwarzschild criterion (Equation \ref{sch}) once the convective zone is developed. The location of the RCB does not depend on the surface optical depth or surface temperature. If we adopt the flux continuity constraint, the actual RCB should be located at a higher level than $\tau_{sc}$ for terrestrial planets (see Figure 1 in \citealt{robinsonAnalyticRadiativeConvectiveModel2012}).

The above discussion applies only to convective instability within the atmosphere. As we discussed in Appendix \ref{app2}, near-surface convection may also occur if $T_{\rm s}^4-S(\tau_{\rm s})>0$ (Equation \ref{surft}). This condition is met, for example, if $D=2$ and $\alpha<D$. The onset of near-surface convection does not require $\tau_{s}>\tau_{sc}$ and should therefore be considered a separate criterion for convective instability on terrestrial planets, distinct from the traditional Schwarzschild criterion.

\section{inequality between two simplest inhomogeneous cases}
\label{app:convex}

Jensen's inequality compares the inhomogeneous case with the homogeneous case. In this appendix, we consider the simplest inhomogeneous cases with only two data points in each case. Let the mean value be $x_0$, and let $X = \{x_0-a, x_0+a\}$ and $Y = \{x_0-b, x_0+b\}$, where $a \ge b \ge 0$. In this case, $X$ is more inhomogeneous than $Y$. We can prove the following proposition.

\textit{Proposition II.} For any convex function $f(x)$, if $a\ge b \ge 0$, then
\begin{equation}
f(x+a)+f(x-a)\geq f(x+b)+f(x-b). \label{2var}
\end{equation}

\textit{Proof.} Since $f(x)$ is convex, for all $x_1, x_2 \in \mathbb{R}$ and for all $\lambda \in [0,1]$, we have
\begin{equation}
\lambda f(x_1) + (1-\lambda) f(x_2) \geq f(\lambda x_1 + (1-\lambda) x_2) .
\end{equation}
Let $\lambda=b/a,~x_1=x-a,~x_2=x$, we have
\begin{equation}
\frac{b}{a} f(x-a) + \left(1-\frac{b}{a}\right) f(x) \geq f(x-b) .\label{eq:x-a}
\end{equation}
Likewise, let $x_1=x+a,~x_2=x$, we have
\begin{equation}
\frac{b}{a} f(x+a) + \left(1-\frac{b}{a}\right) f(x) \geq f(x+b) .\label{eq:x+a}
\end{equation}
Also, let $\lambda=1/2,~x_1=x-a,~x_2=x+a$, we have
\begin{equation}
f(x-a) + f(x+a) \ge 2f(x) .\label{eq:x}
\end{equation}
Thus we can sum Equations $\ref{eq:x-a}+\ref{eq:x+a}+(1-b/a)\ref{eq:x}$ and eliminate $f(x)$ to prove the proposition:
\begin{equation}
f(x+a)+f(x-a)\ge f(x+b)+f(x-b).
\end{equation}

An alternative proof can also be provided here.

\textit{Proof.} Since $f(x)$ is convex, for $a \ge b \ge 0$, we have:
\begin{equation}
f^\prime(x+a+b)\geq f^\prime(x),
\end{equation}
where $f^\prime(x)$ is the first derivative of $f(x)$. Then:
\begin{equation}
\begin{split}
f(x+a)-f(x+b) &= \int_{x+b}^{x+a} f^\prime(t) dt \\
&=\int_{x-a}^{x-b} f^\prime(t+a+b) dt \\
&\geq \int_{x-a}^{x-b} f^\prime(t) dt \\
&= f(x-b)-f(x-a).
\end{split}
\end{equation}
Rearrange to obtain:
\begin{equation}
f(x+a)+f(x-a)\ge f(x+b)+f(x-b).
\end{equation}

Thus, the case with a larger inhomogeneity has a larger mean value if the function is convex. For a concave function, the case with a larger inhomogeneity has a smaller mean value.

% \section{Appendix information}
% \section{IAU recommendations for nominal units \label{nominal}}

%\bibliography{ref}{}

\begin{thebibliography}{}
\expandafter\ifx\csname natexlab\endcsname\relax\def\natexlab#1{#1}\fi
\providecommand{\url}[1]{\href{#1}{#1}}
\providecommand{\dodoi}[1]{doi:~\href{http://doi.org/#1}{\nolinkurl{#1}}}
\providecommand{\doeprint}[1]{\href{http://ascl.net/#1}{\nolinkurl{http://ascl.net/#1}}}
\providecommand{\doarXiv}[1]{\href{https://arxiv.org/abs/#1}{\nolinkurl{https://arxiv.org/abs/#1}}}

\bibitem[{Ag{\'u}ndez {et~al.}(2014)Ag{\'u}ndez, Parmentier, Venot, Hersant, \&
  Selsis}]{agundezPseudo2DChemical2014}
Ag{\'u}ndez, M., Parmentier, V., Venot, O., Hersant, F., \& Selsis, F. 2014,
  Astronomy \& Astrophysics, 564, A73, \dodoi{10.1051/0004-6361/201322895}

\bibitem[{Apai {et~al.}(2017)Apai, Karalidi, Marley, Yang, Flateau, Metchev,
  Cowan, Buenzli, Burgasser, \& Radigan}]{apaiZonesSpotsPlanetaryscale2017}
Apai, D., Karalidi, T., Marley, M.~S., {et~al.} 2017, Science, 357, 683

\bibitem[{Armour {et~al.}(2013)Armour, Bitz, \&
  Roe}]{armourTimevaryingClimateSensitivity2013}
Armour, K.~C., Bitz, C.~M., \& Roe, G.~H. 2013, Journal of Climate, 26, 4518

\bibitem[{Armstrong(1968)}]{armstrongTheoryDiffusivityFactor1968}
Armstrong, B. 1968, Journal of Quantitative Spectroscopy and Radiative
  Transfer, 8, 1577, \dodoi{10.1016/0022-4073(68)90052-6}

\bibitem[{Artigau {et~al.}(2009)Artigau, Bouchard, Doyon, \&
  Lafreni{\`e}re}]{artigauPhotometricVariabilityT22009}
Artigau, {\'E}., Bouchard, S., Doyon, R., \& Lafreni{\`e}re, D. 2009, The
  Astrophysical Journal, 701, 1534, \dodoi{10.1088/0004-637X/701/2/1534}

\bibitem[{Biller {et~al.}(2015)Biller, Vos, Bonavita, Buenzli, Baxter,
  Crossfield, Allers, Liu, Bonnefoy, \&
  Deacon}]{billerVariabilityYoungTransition2015}
Biller, B.~A., Vos, J., Bonavita, M., {et~al.} 2015, The Astrophysical Journal
  Letters, 813, L23

\bibitem[{Budaj {et~al.}(2012)Budaj, Hubeny, \&
  Burrows}]{budajDayNightSide2012}
Budaj, J., Hubeny, I., \& Burrows, A. 2012, Astronomy \& Astrophysics, 537,
  A115, \dodoi{10.1051/0004-6361/201117975}

\bibitem[{Drummond {et~al.}(2020)Drummond, Hebrard, Mayne, Venot, Ridgway,
  Changeat, Tsai, Manners, Tremblin, Abraham, Sing, \&
  Kohary}]{drummondImplicationsThreedimensionalChemical2020}
Drummond, B., Hebrard, E., Mayne, N.~J., {et~al.} 2020, Astronomy \&
  Astrophysics, \dodoi{10.1051/0004-6361/201937153}

\bibitem[{Fauchez {et~al.}(2014)Fauchez, Cornet, Szczap, Dubuisson, \&
  Rosambert}]{fauchezImpactCirrusClouds2014}
Fauchez, T., Cornet, C., Szczap, F., Dubuisson, P., \& Rosambert, T. 2014,
  Atmospheric Chemistry and Physics, 14, 5599, \dodoi{10.5194/acp-14-5599-2014}

\bibitem[{Feinstein {et~al.}(2022)Feinstein, Radica, Welbanks, Murray, Ohno,
  Coulombe, Espinoza, Bean, Teske, Benneke, Line, Rustamkulov, Saba, Tsiaras,
  Barstow, Fortney, Gao, Knutson, MacDonald, {Mikal-Evans}, Rackham, Taylor,
  Parmentier, Batalha, {Berta-Thompson}, Carter, Changeat, Santos, Gibson,
  Goyal, Kreidberg, {L{\'o}pez-Morales}, Lothringer, Miguel, Molaverdikhani,
  Moran, Morello, Mukherjee, Sing, Stevenson, Wakeford, Ahrer, Alam, Alderson,
  Allen, Batalha, Bell, Blecic, Brande, Caceres, Casewell, Chubb, Crossfield,
  Crouzet, Cubillos, Decin, D{\'e}sert, Harrington, Heng, Henning, Iro,
  Kempton, Kendrew, Kirk, Krick, Lagage, Lendl, Mancini, Mansfield, May, Mayne,
  Nikolov, Palle, dit {de la Roche}, Piaulet, Powell, Redfield, Rogers, Roman,
  Roy, Nixon, Schlawin, Tan, Tremblin, Turner, Venot, Waalkes, Wheatley, \&
  Zhang}]{feinsteinEarlyReleaseScience2022}
Feinstein, A.~D., Radica, M., Welbanks, L., {et~al.} 2022, Early {{Release
  Science}} of the Exoplanet {{WASP-39b}} with {{JWST NIRISS}},  {arXiv},
  \dodoi{10.48550/arXiv.2211.10493}

\bibitem[{Fortney {et~al.}(2008)Fortney, Lodders, Marley, \&
  Freedman}]{fortneyUnifiedTheoryAtmospheres2008}
Fortney, J.~J., Lodders, K., Marley, M.~S., \& Freedman, R.~S. 2008, The
  Astrophysical Journal, 678, 1419, \dodoi{10.1086/528370}

\bibitem[{Fortney {et~al.}(2007)Fortney, Marley, \&
  Barnes}]{fortneyPlanetaryRadiiFive2007}
Fortney, J.~J., Marley, M.~S., \& Barnes, J.~W. 2007, The Astrophysical
  Journal, 659, 1661, \dodoi{10.1086/512120}

\bibitem[{Freedman {et~al.}(2014)Freedman, {Lustig-Yaeger}, Fortney, Lupu,
  Marley, \& Lodders}]{freedmanGaseousMeanOpacities2014}
Freedman, R.~S., {Lustig-Yaeger}, J., Fortney, J.~J., {et~al.} 2014, The
  Astrophysical Journal Supplement Series, 214, 25,
  \dodoi{10.1088/0067-0049/214/2/25}

\bibitem[{Ge {et~al.}(2019)Ge, Zhang, Fletcher, Orton, Sinclair, Fernandes,
  Momary, Kasaba, Sato, \& Fujiyoshi}]{geRotationalLightCurves2019}
Ge, H., Zhang, X., Fletcher, L.~N., {et~al.} 2019, The Astronomical Journal,
  157, 89

\bibitem[{Gilli {et~al.}(2021)Gilli, Navarro, Lebonnois, Quirino, Silva,
  Stolzenbach, Lef{\`e}vre, \& Schubert}]{gilliVenusUpperAtmosphere2021}
Gilli, G., Navarro, T., Lebonnois, S., {et~al.} 2021, Icarus, 366, 114432,
  \dodoi{10.1016/j.icarus.2021.114432}

\bibitem[{Ginzburg \& Sari(2015)}]{ginzburgHotJupiterInflationDue2015}
Ginzburg, S., \& Sari, R. 2015, The Astrophysical Journal, 803, 111,
  \dodoi{10.1088/0004-637X/803/2/111}

\bibitem[{Ginzburg \& Sari(2016)}]{ginzburgExtendedHeatDeposition2016}
---. 2016, The Astrophysical Journal, 819, 116,
  \dodoi{10.3847/0004-637X/819/2/116}

\bibitem[{Gladstone \& Young(2019)}]{gladstoneNewHorizonsObservations2019}
Gladstone, G.~R., \& Young, L.~A. 2019, Annual Review of Earth and Planetary
  Sciences, 47, 119, \dodoi{10.1146/annurev-earth-053018-060128}

\bibitem[{Goody \& Yung(1995)}]{goodyAtmosphericRadiationTheoretical1995}
Goody, R.~M., \& Yung, Y.~L. 1995, Atmospheric {{Radiation}}: {{Theoretical
  Basis}} ({Oxford University Press})

\bibitem[{Guillot(2010)}]{guillotRadiativeEquilibriumIrradiated2010}
Guillot, T. 2010, Astronomy \& Astrophysics, 520, A27

\bibitem[{Guillot {et~al.}(1996)Guillot, Burrows, Hubbard, Lunine, \&
  Saumon}]{guillotGiantPlanetsSmall1996}
Guillot, T., Burrows, A., Hubbard, W.~B., Lunine, J.~I., \& Saumon, D. 1996,
  The Astrophysical Journal Letters, 459, L35, \dodoi{10.1086/309935}

\bibitem[{Guillot \& Showman(2002)}]{guillotEvolution51Pegasus2002}
Guillot, T., \& Showman, A.~P. 2002, Astronomy \& Astrophysics, 385, 156

\bibitem[{Guillot {et~al.}(2004)Guillot, Stevenson, Hubbard, \&
  Saumon}]{guillotInteriorJupiter2004}
Guillot, T., Stevenson, D.~J., Hubbard, W.~B., \& Saumon, D. 2004, in Jupiter:
  {{The Planet}}, {{Satellites}} and {{Magnetosphere}}, ed. F.~Bagenal, T.~E.
  Dowling, \& W.~B. McKinnon ({Cambridge Univ. Press}), 35--57

\bibitem[{Hansen(2008)}]{hansenAbsorptionRedistributionEnergy2008}
Hansen, B.~M. 2008, The Astrophysical Journal Supplement Series, 179, 484

\bibitem[{Heng \& Workman(2014)}]{hengAnalyticalModelsExoplanetary2014}
Heng, K., \& Workman, J. 2014, The Astrophysical Journal Supplement Series,
  213, 27, \dodoi{10.1088/0067-0049/213/2/27}

\bibitem[{H{\"o}lder(1889)}]{holder_1889}
H{\"o}lder, O. 1889, 1889, 38, \dodoi{JFM 21.0260.07}

\bibitem[{Hu {et~al.}(2015)Hu, Seager, \& Yung}]{huHELIUMATMOSPHERESWARM2015}
Hu, R., Seager, S., \& Yung, Y.~L. 2015, The Astrophysical Journal, 807, 8,
  \dodoi{10.1088/0004-637X/807/1/8}

\bibitem[{Hubbard(1977)}]{hubbardJovianSurfaceCondition1977}
Hubbard, W.~B. 1977, Icarus, 30, 305, \dodoi{10.1016/0019-1035(77)90164-6}

\bibitem[{Hubeny {et~al.}(2003)Hubeny, Burrows, \&
  Sudarsky}]{hubenyPossibleBifurcationAtmospheres2003}
Hubeny, I., Burrows, A., \& Sudarsky, D. 2003, The Astrophysical Journal, 594,
  1011, \dodoi{10.1086/377080}

\bibitem[{Ingersoll {et~al.}(1975)Ingersoll, Muench, Neugebauer, Diner, Orton,
  Schupler, Schroeder, Chase, Ruiz, \&
  Trafton}]{ingersollPioneer11Infrared1975}
Ingersoll, A.~P., Muench, G., Neugebauer, G., {et~al.} 1975, Science, 188, 472,
  \dodoi{10.1126/science.188.4187.472}

\bibitem[{Jensen(1906)}]{jensenFonctionsConvexesInegalites1906}
Jensen, J. L. W.~V. 1906, Acta Mathematica, 30, 175, \dodoi{10.1007/BF02418571}

\bibitem[{Kasting {et~al.}(1993)Kasting, Whitmire, \&
  Reynolds}]{kastingHabitableZonesMain1993}
Kasting, J.~F., Whitmire, D.~P., \& Reynolds, R.~T. 1993, Icarus, 101, 108,
  \dodoi{10.1006/icar.1993.1010}

\bibitem[{Koll \& Abbot(2016)}]{kollTemperatureStructureAtmospheric2016}
Koll, D.~D., \& Abbot, D.~S. 2016, The Astrophysical Journal, 825, 99

\bibitem[{Koll \& Cronin(2018)}]{kollEarthOutgoingLongwave2018}
Koll, D. D.~B., \& Cronin, T.~W. 2018, Proceedings of the National Academy of
  Science, 115, 10293, \dodoi{10.1073/pnas.1809868115}

\bibitem[{Kopparapu {et~al.}(2013)Kopparapu, Ramirez, Kasting, Eymet, Robinson,
  Mahadevan, Terrien, {Domagal-Goldman}, Meadows, \&
  Deshpande}]{kopparapuHABITABLEZONESMAINSEQUENCE2013}
Kopparapu, R.~K., Ramirez, R., Kasting, J.~F., {et~al.} 2013, The Astrophysical
  Journal, 765, 131, \dodoi{10.1088/0004-637X/765/2/131}

\bibitem[{Larson {et~al.}(2001)Larson, Wood, Field, Golaz, Haar, \&
  Cotton}]{larsonSystematicBiasesMicrophysics2001}
Larson, V.~E., Wood, R., Field, P.~R., {et~al.} 2001, Journal of the
  Atmospheric Sciences, 58, 1117,
  \dodoi{10.1175/1520-0469(2001)058<1117:SBITMA>2.0.CO;2}

\bibitem[{Leconte {et~al.}(2013)Leconte, Forget, Charnay, Wordsworth, Selsis,
  Millour, \& Spiga}]{leconte3DClimateModeling2013}
Leconte, J., Forget, F., Charnay, B., {et~al.} 2013, Astronomy \& Astrophysics,
  554, A69, \dodoi{10.1051/0004-6361/201321042}

\bibitem[{Lee {et~al.}(2023)Lee, Tsai, Hammond, \&
  Tan}]{leeMiniChemicalSchemeNet2023}
Lee, E. K.~H., Tsai, S.-M., Hammond, M., \& Tan, X. 2023, Astronomy \&
  Astrophysics, 672, A110, \dodoi{10.1051/0004-6361/202245473}

\bibitem[{Liao \& Berg(2019)}]{liaoSharpeningJensenInequality2019}
Liao, J.~G., \& Berg, A. 2019, The American Statistician, 73, 278,
  \dodoi{10.1080/00031305.2017.1419145}

\bibitem[{Liou(2002)}]{liouIntroductionAtmosphericRadiation2002}
Liou, K.~N. 2002, An {{Introduction}} to {{Atmospheric Radiation}} ({Elsevier})

\bibitem[{Lohmann(2020)}]{lohmannTemperaturesEnergyBalance2020}
Lohmann, G. 2020, Earth System Dynamics, 11, 1195,
  \dodoi{10.5194/esd-11-1195-2020}

\bibitem[{Malsky {et~al.}(2022)Malsky, Rogers, Kempton, \&
  Marounina}]{malskyHeliumenhancedPlanetsUpper2022}
Malsky, I., Rogers, L., Kempton, E. M.-R., \& Marounina, N. 2022, Nature
  Astronomy, 1, \dodoi{10.1038/s41550-022-01823-8}

\bibitem[{Malsky \& Rogers(2020)}]{malskyCoupledThermalCompositional2020}
Malsky, I., \& Rogers, L.~A. 2020, arXiv preprint arXiv:2002.06466.
\newblock \doeprint{2002.06466}

\bibitem[{Manabe \& Strickler(1964)}]{manabeThermalEquilibriumAtmosphere1964}
Manabe, S., \& Strickler, R.~F. 1964, Journal of the Atmospheric Sciences, 21,
  361

\bibitem[{Marley {et~al.}(1996)Marley, Saumon, Guillot, Freedman, Hubbard,
  Burrows, \& Lunine}]{marleyAtmosphericEvolutionarySpectral1996a}
Marley, M.~S., Saumon, D., Guillot, T., {et~al.} 1996, Science, 272, 1919,
  \dodoi{10.1126/science.272.5270.1919}

\bibitem[{McKay {et~al.}(1989)McKay, Pollack, \&
  Courtin}]{mckayThermalStructureTitan1989}
McKay, C.~P., Pollack, J.~B., \& Courtin, R. 1989, Icarus, 80, 23,
  \dodoi{10.1016/0019-1035(89)90160-7}

\bibitem[{Moses {et~al.}(2021)Moses, Tremblin, Venot, \&
  Miguel}]{mosesChemicalVariationAltitude2021}
Moses, J.~I., Tremblin, P., Venot, O., \& Miguel, Y. 2021, Experimental
  Astronomy, \dodoi{10.1007/s10686-021-09749-1}

\bibitem[{Parmentier \& Guillot(2014)}]{parmentierNongreyAnalyticalModel2014}
Parmentier, V., \& Guillot, T. 2014, Astronomy \& Astrophysics, 562, A133

\bibitem[{Parmentier {et~al.}(2015)Parmentier, Guillot, Fortney, \&
  Marley}]{parmentierNongreyAnalyticalModel2015}
Parmentier, V., Guillot, T., Fortney, J.~J., \& Marley, M.~S. 2015, Astronomy
  \& Astrophysics, 574, A35

\bibitem[{Parmentier {et~al.}(2013)Parmentier, Showman, \&
  Lian}]{parmentier3DMixingHot2013}
Parmentier, V., Showman, A.~P., \& Lian, Y. 2013, Astronomy \& Astrophysics,
  558, A91

\bibitem[{Pierrehumbert(2010)}]{pierrehumbertPrinciplesPlanetaryClimate2010}
Pierrehumbert, R.~T. 2010, Principles of {{Planetary Climate}} ({Cambridge
  University Press})

\bibitem[{Powell {et~al.}(2019)Powell, Louden, Kreidberg, Zhang, Gao, \&
  Parmentier}]{powellTransitSignaturesInhomogeneous2019}
Powell, D., Louden, T., Kreidberg, L., {et~al.} 2019, The Astrophysical
  Journal, 887, 170

\bibitem[{Rauscher \&
  Showman(2014)}]{rauscherINFLUENCEDIFFERENTIALIRRADIATION2014}
Rauscher, E., \& Showman, A.~P. 2014, The Astrophysical Journal, 784, 160,
  \dodoi{10.1088/0004-637X/784/2/160}

\bibitem[{Robinson \&
  Catling(2012)}]{robinsonAnalyticRadiativeConvectiveModel2012}
Robinson, T.~D., \& Catling, D.~C. 2012, The Astrophysical Journal, 757, 104,
  \dodoi{10.1088/0004-637X/757/1/104}

\bibitem[{Rodgers \& Walshaw(1966)}]{rodgersComputationInfraredCooling1966}
Rodgers, C.~D., \& Walshaw, C.~D. 1966, Quarterly Journal of the Royal
  Meteorological Society, 92, 67, \dodoi{10.1002/qj.49709239107}

\bibitem[{Rogers(1888)}]{rogersExtensionCertainTheorem1888}
Rogers, L.~J. 1888, Messenger of Math., 17, 145, \dodoi{JFM 20.0254.02}

\bibitem[{Shao {et~al.}(2022)Shao, Zhang, Mendon{\c c}a, \&
  Encrenaz}]{shaoLocaltimeDependenceChemical2022}
Shao, W.~D., Zhang, X., Mendon{\c c}a, J., \& Encrenaz, T. 2022, The Planetary
  Science Journal, 3, 3, \dodoi{10.3847/PSJ/ac3bd3}

\bibitem[{Showman(2007)}]{showmanNumericalSimulationsForced2007}
Showman, A.~P. 2007, Journal of Atmospheric Sciences, 64, 3132

\bibitem[{Sobel {et~al.}(2001)Sobel, Nilsson, \&
  Polvani}]{sobelWeakTemperatureGradient2001}
Sobel, A.~H., Nilsson, J., \& Polvani, L.~M. 2001, Journal of Atmospheric
  Sciences, 58, 3650, \dodoi{10.1175/1520-0469(2001)058<3650:TWTGAA>2.0.CO;2}

\bibitem[{Spiegel \& Burrows(2013)}]{spiegelThermalProcessesGoverning2013}
Spiegel, D.~S., \& Burrows, A. 2013, The Astrophysical Journal, 772, 76,
  \dodoi{10.1088/0004-637X/772/1/76}

\bibitem[{Strobel \& Zhu(2017)}]{strobelComparativePlanetaryNitrogen2017}
Strobel, D.~F., \& Zhu, X. 2017, Icarus, 291, 55,
  \dodoi{10.1016/j.icarus.2017.03.013}

\bibitem[{Tan(2022)}]{tanWeakSeasonalityTemperate2022}
Tan, X. 2022, The Astrophysical Journal, 926, 202,
  \dodoi{10.3847/1538-4357/ac4d8a}

\bibitem[{Tan \&
  Showman(2021{\natexlab{a}})}]{tanAtmosphericCirculationBrown2021a}
Tan, X., \& Showman, A.~P. 2021{\natexlab{a}}, Monthly Notices of the Royal
  Astronomical Society, 502, 678, \dodoi{10.1093/mnras/stab060}

\bibitem[{Tan \&
  Showman(2021{\natexlab{b}})}]{tanAtmosphericCirculationBrown2021}
---. 2021{\natexlab{b}}, Monthly Notices of the Royal Astronomical Society,
  502, 2198, \dodoi{10.1093/mnras/stab097}

\bibitem[{Turbet {et~al.}(2021)Turbet, Bolmont, Chaverot, Ehrenreich, Leconte,
  \& Marcq}]{turbetDayNightCloud2021}
Turbet, M., Bolmont, E., Chaverot, G., {et~al.} 2021, Nature, 598, 276,
  \dodoi{10.1038/s41586-021-03873-w}

\bibitem[{Weaver \& Ramanathan(1995)}]{weaverDeductionsSimpleClimate1995}
Weaver, C.~P., \& Ramanathan, V. 1995, Journal of Geophysical Research, 100,
  11585, \dodoi{10.1029/95JD00770}

\bibitem[{Wordsworth(2015)}]{wordsworthAtmosphericHeatRedistribution2015}
Wordsworth, R. 2015, The Astrophysical Journal, 806, 180

\bibitem[{Yang {et~al.}(2013)Yang, Cowan, \&
  Abbot}]{yangStabilizingCloudFeedback2013}
Yang, J., Cowan, N.~B., \& Abbot, D.~S. 2013, The Astrophysical Journal
  Letters, 771, L45

\bibitem[{Zhang(2020)}]{zhangAtmosphericRegimesTrends2020}
Zhang, X. 2020, Research in Astronomy and Astrophysics, 20, 099,
  \dodoi{10.1088/1674-4527/20/7/99}

\bibitem[{Zhang(2023)}]{zhangInhomogeneityEffectII2023}
---. 2023, The Astrophysical Journal, in press

\bibitem[{Zhang {et~al.}(2023)Zhang, Li, Ge, \&
  Le}]{zhangInhomogeneityEffectIII2023}
Zhang, X., Li, C., Ge, H., \& Le, T. 2023, The Astrophysical Journal, in press

\bibitem[{Zhang \& Showman(2014)}]{zhangAtmosphericCirculationBrown2014}
Zhang, X., \& Showman, A.~P. 2014, The Astrophysical Journal Letters, 788, L6

\bibitem[{Zhang \& Showman(2017)}]{zhangEffectsBulkComposition2017}
---. 2017, The Astrophysical Journal, 836, 73

\bibitem[{Zhang \&
  Showman(2018{\natexlab{a}})}]{zhangGlobalmeanVerticalTracer2018}
---. 2018{\natexlab{a}}, The Astrophysical Journal, 866, 1

\bibitem[{Zhang \&
  Showman(2018{\natexlab{b}})}]{zhangGlobalmeanVerticalTracer2018a}
---. 2018{\natexlab{b}}, The Astrophysical Journal, 866, 2

\bibitem[{Zhang {et~al.}(2017)Zhang, Strobel, \&
  Imanaka}]{zhangHazeHeatsPluto2017}
Zhang, X., Strobel, D.~F., \& Imanaka, H. 2017, Nature, 551, 352,
  \dodoi{10.1038/nature24465}

\bibitem[{Zhang {et~al.}(2013)Zhang, West, Banfield, \&
  Yung}]{zhangStratosphericAerosolsJupiter2013}
Zhang, X., West, R.~A., Banfield, D., \& Yung, Y.~L. 2013, Icarus, 226, 159,
  \dodoi{10.1016/j.icarus.2013.05.020}

\bibitem[{Zhao \& Shi(2013)}]{zhaoAccurateApproximationDiffusivity2013}
Zhao, J.-Q., \& Shi, G.-Y. 2013, Infrared Physics \& Technology, 56, 21,
  \dodoi{10.1016/j.infrared.2012.09.003}

\bibitem[{Zhou {et~al.}(2016)Zhou, Apai, Schneider, Marley, \&
  Showman}]{zhouDiscoveryRotationalModulations2016}
Zhou, Y., Apai, D., Schneider, G.~H., Marley, M.~S., \& Showman, A.~P. 2016,
  The Astrophysical Journal, 818, 176

\end{thebibliography}
%\bibliographystyle{aasjournal}

%\iffalse

%\fi

%% Include this line if you are using the \added, \replaced, \deleted
%% commands to see a summary list of all changes at the end of the article.
%\listofchanges

\end{document}